\documentclass[a4paper,11pt]{article}
\pdfoutput=1
\usepackage{jheppub} % see the JHEP-author-manual
\usepackage[utf8]{inputenc}
\usepackage[T1]{fontenc} % if needed
\usepackage{slashed}
\usepackage{xspace}
\usepackage{dsfont}
\usepackage{amsmath}
\usepackage[normalem]{ulem}
\usepackage{booktabs}
\usepackage{pbox}

\title{Completing the scalar and fermionic Universal One-Loop Effective Action}
\author{Michael Kr\"amer,}
\author{Benjamin Summ}
\author{and Alexander Voigt}

\affiliation{Institute for Theoretical Particle Physics and Cosmology,\\
  RWTH Aachen University, 52074 Aachen, Germany}
\emailAdd{mkraemer@physik.rwth-aachen.de}
\emailAdd{benjamin.summ@rwth-aachen.de}
\emailAdd{alexander.voigt@physik.rwth-aachen.de}

\abstract{We extend the known Universal One-Loop Effective Action
  (UOLEA) by all operators which involve scalars and fermions, not including  contributions arising from open covariant derivatives. Our generic analytic expressions for the one-loop Wilson coefficients of effective operators up to dimension six allow for an application of the UOLEA to a broader class of UV-complete models. We apply our generic results to various effective theories of
  supersymmetric models, where different supersymmetric particles are integrated
  out at a high mass scale.}
  
\keywords{UOLEA, EFT, MSSM}

\arxivnumber{}

\hypersetup{
  pdftitle    = {Completing the scalar and fermionic Universal One-Loop Effective Action},
  pdfauthor   = {Michael Kr\"amer, Benjamin Summ, Alexander Voigt},
  pdfkeywords = {UOLEA, EFT, MSSM}
}

\newcommand{\secref}[1]{section~\ref{#1}}

\newcommand{\appref}[1]{appendix~\ref{#1}}
\newcommand{\MSbar}{\ensuremath{\overline{\text{MS}}}\xspace}
\newcommand{\DRbar}{\ensuremath{\overline{\text{DR}}}\xspace}
\newcommand{\DRbarPrime}{\ensuremath{\overline{\text{DR}}'}\xspace}

\newcommand{\GeV}{\ensuremath{\;\text{GeV}}}

\newcommand{\EFT}{\ensuremath{\text{EFT}}\xspace}
\newcommand{\UV}{\ensuremath{\text{UV}}\xspace}
\newcommand{\tree}{\ensuremath{\text{tree}}\xspace}
\newcommand{\FS}{\texttt{FlexibleSUSY}\@\xspace}

\newcommand{\SARAH}{\texttt{SARAH}\@\xspace}
\newcommand{\CoDEx}{\texttt{CoDEx}\@\xspace}
\newcommand{\Lag}{\mathcal{L}}
\newcommand{\Lagint}{\Lag_{\text{UV,int}}}

\newcommand{\diag}{\mathop{\rm diag}}
\newcommand{\eps}{\epsilon}
\newcommand{\epsIR}{\eps_{\text{IR}}}
\newcommand{\epsUV}{\eps_{\text{UV}}}
\newcommand{\epstensor}{\varepsilon}
\newcommand{\SM}{\ensuremath{\text{SM}}\xspace}
\newcommand{\SMEFT}{\ensuremath{\text{SMEFT}}\xspace}
\newcommand{\MSSM}{\ensuremath{\text{MSSM}}\xspace}

\newcommand{\hc}{\text{h.c.}}
\newcommand{\Loop}{\ensuremath{\ell}}

\newcommand{\re}{\ensuremath{\Re\mathfrak{e}}}
\newcommand{\rd}{\ensuremath{\mathrm{d}}}
\DeclareMathOperator{\tr}{tr}

\newcommand{\cc}{\mathcal{C}}
\newcommand{\ccfield}[1]{#1^C}
\newcommand{\classicfield}[1]{#1_{\text{cl}}}
\newcommand{\ZI}{\tilde{\mathcal{I}}}
\newcommand{\order}[1]{\ensuremath{\mathcal{O}(#1)}}
\newcommand{\fourdim}[1]{\ensuremath{\mathring{#1}}}
\newcommand{\ddim}[1]{\ensuremath{#1}}
\newcommand{\epsdim}[1]{\ensuremath{\breve{#1}}}
\newcommand{\fluct}{\mathcal{Q}} % fluctuation operator
\newcommand{\fluctS}{\mathcal{Q}_{\text{S}}} % scalar fluctuation operator
\DeclareMathOperator{\Tr}{Tr}

\newcommand{\slep}{\tilde{l}}
\newcommand{\sel}{\tilde{e}}
\newcommand{\sneu}{\tilde{\nu}}
\newcommand{\sq}{\tilde{q}}
\newcommand{\su}{\tilde{u}}
\newcommand{\sd}{\tilde{d}}

\newcommand{\st}[1]{\tilde{t}_{#1}}

\newcommand{\gluino}[1]{\tilde{g}^{#1}}
\newcommand{\gluinocl}{\classicfield{\tilde{g}}}
\newcommand{\phicl}{\classicfield{\phi}}
\newcommand{\Phicl}{\classicfield{\Phi}}

\newcommand{\mstL}{m^2_{\sq}}
\newcommand{\mstR}{m^2_{\su}}

\allowdisplaybreaks

\begin{document}

\begin{flushright}
\footnotesize
TTK--19--31 \\
\footnotesize
P3H--19--026
\end{flushright}
\maketitle
\flushbottom

\section{Introduction}

% SM is complete, but not a full description of nature
%
With the discovery of the Higgs boson at the Large Hadron Collider (LHC) \cite{Aad:2012tfa,Chatrchyan:2012xdj}, the Standard Model of
Particle Physics (SM) is formally complete. While existing deviations between some SM predictions and experiment, such as for the anomalous
magnetic moment of the muon (see for example \cite{Bennett:2006fi, Jegerlehner:2018zrj}), are not conclusive, the SM is not a complete description of nature as it neither accounts for  astrophysical phenomena such as dark matter, nor does it incorporate gravity.
%

% new particles may be heavy:
%
Searches for physics beyond the SM have not been successful thus far. Exclusion limits for new particles introduced by SM extensions often exceed the TeV scale. These results suggest that new physics either interacts weakly with the SM, or that the masses of new particles are significantly above the electroweak scale.  A well-known example is the Minimal
Supersymmetric Standard Model (MSSM)~\cite{Haber:1984rc}, which requires at least
TeV-scale stops in order to correctly predict the mass of the SM-like
Higgs boson of about $125\GeV$, see for example
\cite{Allanach:2018fif,Bahl:2018zmf}.
The construction and phenomenological analysis of new physics models with heavy particles is therefore a suitable path to develop viable theories beyond the SM that are consistent with experimental results.

% construct EFTs:
%
The observables predicted in models with large mass hierarchies,
however, usually suffer from large logarithmic quantum corrections,
which should be resummed in order to obtain precise predictions.
Effective Field Theories (EFTs) are a well-suited tool to resum these
large logarithmic corrections.  Conventional matching procedures using
Feynman diagrams, however, are often cumbersome, in particular if the
new physics model contains many new heavy particles and/or complicated
interactions.  The Universal One-Loop Effective Action (UOLEA)
\cite{Drozd:2015rsp,Ellis:2017jns,Summ:2018oko}, which has been
developed using functional methods
\cite{Gaillard:1985uh,Cheyette:1987qz,Haba:2011vi,Henning:2014wua,Henning:2016lyp,Ellis:2016enq,Fuentes-Martin:2016uol,Zhang:2016pja},
is a very promising tool to overcome these difficulties.  It represents a
generic one-loop expression for the Wilson coefficients of an effective
Lagrangian for a given ultra-violet (UV) model with a large mass
hierarchy.  Compared to the conventional matching using Feynman
diagrams, the calculation of the Wilson coefficients with the UOLEA is
straightforward, as it is expressed directly in terms of derivatives
of the UV Lagrangian w.r.t.\ the fields and simple rational functions.
In particular, no loop integration is necessary and spurious infrared
(IR) divergences are absent by construction.
%
% status:
%
To date, however, the UOLEA is not completely known: Only
contributions from scalar particles \cite{Drozd:2015rsp,Ellis:2017jns}
as well as conversion terms between dimensional regularization and dimensional reduction \cite{Summ:2018oko} have been
calculated at the generic one-loop level up to dimension 6.  Whereas
some contributions from fermion loops can be calculated using these
results by squaring the fermionic trace, this treatment is incomplete
when the couplings depend on gamma matrices. Furthermore,
contributions from loops containing both scalars and fermions as well
as terms with open covariant derivatives are unknown.

% this paper:
%
In this publication we present all one-loop operators of the UOLEA up to
dimension 6 that involve both scalars and fermions in a generic form,
excluding contributions from open covariant derivatives.
Thus, our results go beyond the scope of
\cite{Drozd:2015rsp,Ellis:2017jns} and allow for an application of the
UOLEA to a broader set of new physics models.  We publish our generic
expressions as a Mathematica ancillary file \texttt{UOLEA.m} in the
arXiv submission of this publication.
Due to their generic structure, the expressions are well suited to be
implemented into generic spectrum generators such as \SARAH
\cite{Staub:2009bi,Staub:2010jh,Staub:2012pb, Staub:2013tta} or \FS
\cite{Athron:2014yba,Athron:2017fvs} or EFT codes in the spirit of
\CoDEx \cite{Bakshi:2018ics,DasBakshi:2019vzr}.

This paper is structured as follows: In \secref{sec:calculation} we
present the calculation of the UOLEA involving both scalars and
fermions.  We discuss the results in \secref{sec:results} and apply our generic expressions 
to various EFTs of the SM and the MSSM in \secref{sec:applications}. Our conclusions are presented in \secref{sec:conclusions}, and the appendices collect further formulae and calculational details.

%%%%%%%%%%%%%%%%%%%%%%%%%%%%%%%%%%%%%%%%%%%%%%%%%%

\section{Calculation of the scalar and fermionic UOLEA}
\label{sec:calculation}

\subsection{Functional matching in a scalar theory}
\label{sec: intro}

In this section we briefly review the most important steps in the
functional matching approach at one-loop level in a scalar theory and fix the notation
for the subsequent sections. Most of what is being discussed here is
well-documented in the literature and more details can be found in
\cite{Henning:2014wua,Henning:2016lyp,Fuentes-Martin:2016uol,Zhang:2016pja}. We
consider a generic UV theory that contains heavy real scalar fields,
collectively denoted by $\Phi$, with masses of the order $M$ and light
real scalar fields, denoted by $\phi$, with masses of the order $m$.
We assume that $m/M \ll 1$ such that an EFT expansion in the mass
ratio $m/M$ is valid.  To perform the functional matching the
background field method is used to calculate the generator of
1-light-particle-irreducible (1LPI) Green's functions in the
UV-theory, $\Gamma_{\text{L,UV}}[\phicl]$, and the generator of
1-particle-irreducible (1PI) Green's functions in the EFT,
$\Gamma_{\EFT}[\phicl]$, where $\phicl$ are light background
fields which obey the classical equation of motion. For the determination of these generating functionals beyond
tree-level a regularization scheme must be specified, which is chosen
to be dimensional regularization.\footnote{In principle the results
  obtained in this paper can also be applied to a setting where
  dimensional reduction is used as a regularization scheme, see \cite{Summ:2018oko}.} This
introduces a dependence on the unphysical renormalization scale $\mu$
in both generating functionals, and the matching condition becomes
\begin{align}
  \Gamma_\text{L,UV}[\phicl]=\Gamma_\EFT[\phicl],
  \label{eq:scalar_matching_condition}
\end{align}
which is imposed at the matching scale $\mu$, order by order in perturbation theory. In principle the matching scale can be chosen arbitrarily, however, in order to avoid large logarithms the choice $\mu=M$ is preferred. To calculate
$\Gamma_\text{L,UV}[\phicl]$ one starts from the generating functional
of Green's functions
\begin{align}
Z_\UV[J_\Phi,J_\phi]=\int \mathcal{D}\Phi \mathcal{D}\phi \exp\left \{i \int \rd^d x \, \big[\Lag_\UV[\Phi,\phi]+J_{\Phi}(x) \Phi(x)+J_{\phi}(x) \phi(x) \big]\right\}
\end{align}
with sources $J_\Phi$ and $J_\phi$ and splits both the heavy and the
light fields into background parts $\Phicl$ and $\phicl$, respectively,
and fluctuations $\delta \Phi$ and $\delta \phi$, respectively, as
\begin{align}
\Phi&=\Phicl+\delta \Phi, \\
\phi&=\phicl+\delta \phi.
\end{align}
The background fields are defined to satisfy the classical equations
of motion,
\begin{align}
  \frac{\delta \Lag_\UV}{\delta \Phi}[\Phicl,\phicl]+J_\Phi &= 0, &
  \frac{\delta \Lag_\UV}{\delta \phi}[\Phicl,\phicl]+J_\phi &= 0.
\end{align}
The generating functional of the 1LPI Green's functions of the UV
model, $\Gamma_\text{L,UV}[\phicl]$, is then given by
\begin{align}
\Gamma_\text{L,UV}[\phicl]=-i \log Z_\UV[J_\Phi=0,J_\phi]-\int \rd^d x \, J_\phi(x) \phicl(x),
\end{align}
where $J_\Phi=0$ since we are only interested in Green's functions
with light external particles.  Expanding the Lagrangian together with
the source terms around the background fields yields
\begin{align}
  \Lag_\UV[\Phi,\phi]+J_{\Phi}\Phi+J_{\phi}\phi &=
  \Lag_\UV[\Phicl,\phicl] + J_{\Phi}\Phicl + J_{\phi}\phicl
  -\frac{1}{2}\begin{pmatrix} \delta \Phi^T  && \delta \phi^T  \end{pmatrix} 
  \fluct
  \begin{pmatrix}
    \delta \Phi \\  \delta \phi
  \end{pmatrix} + \cdots ,
  \label{eq:actionExpansion} \\
  \intertext{where the matrix}
  \fluct &\equiv -
    \begin{pmatrix}
      \frac{\delta ^2 \Lag_\UV}{\delta \Phi \delta \Phi}[\Phicl,\phicl] && \frac{\delta ^2 \Lag_\UV}{\delta \Phi \delta \phi}[\Phicl,\phicl] \\
      \frac{\delta ^2 \Lag_\UV}{\delta \phi \delta \Phi}[\Phicl,\phicl] && \frac{\delta ^2 \Lag_\UV}{\delta \phi \delta \phi}[\Phicl,\phicl]
    \end{pmatrix}
\end{align}
is referred to as the fluctuation operator and the dots indicate
higher order terms in the expansion. Through the equations of motion
with $J_\Phi=0$ the heavy background fields can be expressed in terms
of the light ones such that $\Phicl=\Phicl[\phicl]$. In general,
$\Phicl[\phicl]$ is a non-local object and has to be expanded using a
local operator expansion. The one-loop part of
$\Gamma_\text{L,UV}[\phicl]$ is then found to be
\begin{align}
\Gamma^\text{1\Loop}_\text{L,UV}[\phicl]= \frac{i}{2} \log \det \fluct.
\end{align}
The above can be re-written as \cite{Fuentes-Martin:2016uol}
\begin{align}
\Gamma^\text{1\Loop}_\text{L,UV}[\phicl]&=\frac{i}{2} \log \det
\left(\fluct_{11} - \fluct_{12} \fluct_{22}^{-1} \fluct_{21}\right)
+\frac{i}{2}\log \det \fluct_{22}.
\end{align}
Using similar arguments for the Lagrangian of the EFT,
$\Lag_\EFT[\phi]$, which only depends on the light fields, the
generator of 1PI Green's functions in the EFT can be calculated at
one-loop as
\begin{align}
\Gamma^\text{1\Loop}_\EFT[\phicl]=\int \rd^d x \, \Lag_\EFT^\text{1\Loop}[\phicl]+\frac{i}{2} \log \det \left(-\frac{\delta ^2 \Lag_\EFT^\tree}{\delta \phi \delta \phi}[\phicl] \right),
\end{align}
where $\Lag_\EFT^\text{1\Loop}$ is the effective Lagrangian whose
couplings are given by the one-loop heavy or heavy/light field
contributions.  The second term contains one-loop contributions
constructed from the tree-level part of the effective Lagrangian
$\Lag_\EFT^\tree$. The matching condition
\eqref{eq:scalar_matching_condition} then implies
\begin{align}
\label{eq:matchingCond}
\int \rd^d x \, \Lag_\EFT^\text{1\Loop}[\phi] ={}& \frac{i}{2} \log \det
\left(\fluct_{11} - \fluct_{12} \fluct_{22}^{-1} \fluct_{21}\right) +\frac{i}{2} \log \det \fluct_{22} \nonumber \\ 
& -\frac{i}{2} \log \det \left(-\frac{\delta ^2 \Lag_\EFT^\tree}{\delta \phi \delta \phi}[\phicl] \right).
\end{align}
The functional determinants can be calculated using the relation
$\log \det A = \Tr \log A$ and then calculating the trace. This includes a trace in the Hilbert space as constructed
in \cite{Ball:1988xg}. It is convenient to calculate this trace in
position space and insert the identity in terms of a complete set of
momentum eigenstates. The calculation then involves an integral over
the four-momentum, and expansion by regions
\cite{Beneke:1997zp,Jantzen:2011nz} can be applied to the integrals
\cite{Fuentes-Martin:2016uol,Zhang:2016pja}. It can then be shown
\cite{Zhang:2016pja} that
\begin{align}
\Lag_\EFT^\text{1\Loop}[\phi]&=\frac{i}{2} \int \frac{\rd^dq}{(2\pi)^d} \tr \log \left.
\left(\fluct_{11} - \fluct_{12} \fluct_{22}^{-1} \fluct_{21}\right)\right \rvert ^{P\rightarrow P-q} _\text{hard},
\label{eq:scalarres}
\end{align}
where the final result is given by the hard part of the integrals,
i.e.\ the part for which the integrands can be expanded in the region
$|q^2| \sim M^2 \gg m^2$ and where $P_\mu=i D_\mu$ with $D_\mu$ being
the gauge-covariant derivative. In \eqref{eq:scalarres} the trace over
the Hilbert space has already been performed and ``$\tr$'' designates a
trace over all indices. To derive the currently known form of the
purely scalar UOLEA \cite{Drozd:2015rsp,Ellis:2017jns} from
\eqref{eq:scalarres}, one expands the logarithm in a power series,
which is evaluated up to terms giving rise to operators of
mass dimension 6 and calculates the corresponding coefficients arising
from the momentum integral.  In order to keep gauge-invariance
manifest in the resulting $\Lag_\EFT^\text{1\Loop}$ a covariant
derivative expansion~\cite{Gaillard:1985uh,Cheyette:1987qz} is used, where $P^\mu$ is kept as a whole
and not split into a partial derivative and gauge fields.

\subsection{Fermionic contributions to the UOLEA}
\label{sec:calc}

In this section we consider a more general theory which contains both
scalar and fermionic fields and calculate their contributions to the
UOLEA.\footnote{As discussed in \cite{Zhang:2016pja} and
  \secref{sec:results_vectors}, our final expression for the UOLEA can
  also be used in a more general setting, including, for example,
  massive vector fields.} This extends the results provided in
\cite{Ellis:2017jns} by including contributions to the matching from
loops containing both scalars and fermions as well as contributions
from purely fermionic loops. The latter are partially contained in the
results of \cite{Ellis:2017jns} since they can be computed by squaring
the purely fermionic trace. However, in this approach contributions
are missed whenever the interaction terms among fermions contain gamma
matrices. These terms would be classified as terms with open covariant
derivatives in the language used in \cite{Ellis:2017jns}. In our
treatment no assumptions are made about the spin structure of the
fermionic interactions. In principle, the calculation can be performed
using the method of covariant diagrams introduced in
\cite{Zhang:2016pja}, however, the calculation is presented starting
from first principles for the following reason.  There is some freedom
in choosing the degrees of freedom to integrate over in the path
integral.  For complex scalar fields, for example, these can be the
real and imaginary parts of the field.  Alternatively one can choose
the field and its conjugate as independent degrees of freedom.  For
fermions similar choices can be made.  The explicit form of the
fluctuation operator and the transformations necessary to bring the
Gaussian path integral into a form where it can be trivially
performed depend on this choice.  To reduce
the number of these transformations we use a formalism where Dirac and
Majorana fermions are treated together in one multiplet in the
diagonalization step.  Our formalism has the additional advantage,
that the resulting expressions are more compact compared to the case
when Dirac and Majorana fermions are treated separately.  In the
following we will present our formalism in detail and introduce the
notation of the final result.
  
As mentioned above, there is some freedom in the choice of degrees of
freedom to be integrated over. In order to treat real and complex
scalar fields on the same footing one could split all complex fields
into a real part and an imaginary part and perform the calculation
using these as the fundamental fields. However, for scalars it is
often desirable to maintain the complex fields as they might have some
physical interpretation in the effective theory.  We therefore use the
field and its complex conjugate as independent degrees of freedom. Similarly, in order to
treat Dirac and Majorana fermions simultaneously without diagonalizing
the fluctuation operator among these it is convenient to treat any
Dirac fermion and its charge conjugate as independent degrees of
freedom. We collect all light and heavy scalars into the multiplets
$\phi$ and $\Phi$, respectively, and all light and heavy fermions into
the multiplets $\xi$ and $\Xi$, respectively, see table~\ref{table1}.
\begin{table}[tb]
\centering
\begin{tabular}{cll}
  \toprule
  Multiplet & Components & Description \\
  \midrule
  $\Xi$ & $\big(\Omega, \ccfield{\Omega}, \Lambda\big)^T$ & \pbox{20cm}{$\Omega$, $\ccfield{\Omega}$: heavy Dirac fermions \\
  $\Lambda$: heavy Majorana fermions}  \\
  \midrule
  $\Phi$ & $\big(\Sigma, \Sigma^*, \Theta\big)^T$ & \pbox{20cm}{$\Sigma$, $\Sigma^*$: heavy complex scalars \\
  $\Theta$: heavy real scalars}  \\
  \midrule
  $\xi$ & $\big(\omega, \ccfield{\omega}, \lambda\big)^T$ & \pbox{20cm}{$\omega$, $\ccfield{\omega}$: light Dirac fermions \\
  $\lambda$: light Majorana fermions} \\
  \midrule
  $\phi$ & $\big(\sigma, \sigma^*, \theta\big)^T$ & \pbox{20cm}{$\sigma$, $\sigma^*$: light complex scalars \\
  $\theta$: light real scalars} \\
  \bottomrule
\end{tabular}
\caption{Contents of the different multiplets appearing in the calculation.}
\label{table1}
\end{table}
The charge conjugate of the Dirac spinor $\Omega$ is denoted as
$\ccfield{\Omega}=\cc\bar{\Omega}^T$, with $\cc$ being the charge
conjugation matrix.  Similarly, we define for a light Dirac spinor
$\omega$, $\ccfield{\omega}=\cc\bar{\omega}^T$.  With these
definitions we may write the second variation of the Lagrangian as
follows
\begin{align}
\delta^2 \Lag &= \delta^2 \Lag_\text{S} +\frac{1}{2}\delta \Xi ^T \mathbf{\Delta}_\Xi \delta \Xi - \frac{1}{2} \delta \Xi ^T \tilde{\mathbf{X}}_{\Xi \Phi} \delta \Phi+\frac{1}{2} \delta \Phi ^T \tilde{\mathbf{X}}_{\Phi \Xi} \delta \Xi -\frac{1}{2} \delta \Xi ^T \tilde{\mathbf{X}}_{\Xi \phi} \delta \phi+\frac{1}{2} \delta \phi ^T \tilde{\mathbf{X}}_{\phi \Xi} \delta \Xi \nonumber \\
& \quad+\frac{1}{2} \delta \xi ^T \tilde{\mathbf{X}}_{\xi \Xi} \delta \Xi +\frac{1}{2}\delta \Xi ^T \tilde{\mathbf{X}}_{\Xi \xi} \delta \xi+\frac{1}{2}\delta \xi ^T \mathbf{\Delta}_\xi \delta \xi -\frac{1}{2} \delta \xi ^T \tilde{\mathbf{X}}_{\xi \Phi} \delta \Phi +\frac{1}{2} \delta \Phi ^T \tilde{\mathbf{X}}_{\Phi \xi} \delta \xi \nonumber \\ & \quad -\frac{1}{2} \delta \xi ^T \tilde{\mathbf{X}}_{\xi \phi} \delta \phi+\frac{1}{2} \delta \phi ^T \tilde{\mathbf{X}}_{\phi \xi} \delta \xi,
\label{eq:variation-part1}
\end{align}
where the pure scalar part is given by
\begin{align}
\delta^2 \Lag_\text{S}=-\frac{1}{2} \delta \Phi ^T \mathbf{\Delta}_{\Phi} \delta \Phi -\frac{1}{2} \delta \phi ^T \mathbf{\Delta}_{\phi} \delta \phi -\frac{1}{2} \delta \Phi ^T \tilde{\mathbf{X}}_{\Phi \phi} \delta \phi-\frac{1}{2}\delta \phi ^T \tilde{\mathbf{X}}_{\phi \Phi} \delta \Phi.
\label{eq:variation-part2}
\end{align}
In eqs.~\eqref{eq:variation-part1} and \eqref{eq:variation-part2} we
introduced the following abbreviations:
\begin{align}
\mathbf{\Delta}_\Xi &= \begin{pmatrix} X_{\Omega \Omega} && \cc(\slashed{P}_{\ccfield{\Omega}}-M_\Omega+\cc^{-1} X_{\Omega \bar{\Omega}}\cc^{-1}) && X_{\Omega \Lambda} \\
\cc (\slashed{P}_\Omega-M_\Omega+X_{\bar{\Omega} \Omega}) && \cc X_{\bar{\Omega} \bar{\Omega}} \cc^{-1} && \cc X_{\bar{\Omega} \Lambda} \\
X_{\Lambda \Omega} && X_{\Lambda \bar{\Omega}}\cc ^{-1} && \cc (\slashed{P}_\Lambda-M_\Lambda+\cc ^{-1} X_{\Lambda \Lambda})
\end{pmatrix}
\label{eq:initialDeltaXi}, \\ 
\tilde{\mathbf{X}}_{\Xi \Phi} &= \begin{pmatrix}
X_{\Omega \Sigma} && X_{\Omega \Sigma^{*}} && X_{\Omega \Theta} \\ 
\cc X_{\bar{\Omega} \Sigma} && \cc X_{\bar{\Omega} \Sigma ^*} && \cc X_{\bar{\Omega} \Theta} \\ 
X_{\Lambda \Sigma} && X_{\Lambda \Sigma ^*} && X_{\Lambda \Theta}
\end{pmatrix}, \\ 
\tilde{\mathbf{X}}_{\Phi \Xi} &= \begin{pmatrix}
X_{\Sigma \Omega} && X_{\Sigma \bar{\Omega}} \cc ^{-1} && X_{\Sigma \Lambda} \\
X_{\Sigma ^* \Omega} && X_{\Sigma ^* \bar{\Omega}} \cc ^{-1} && X_{\Sigma ^* \Lambda} \\
X_{\Theta \Omega} && X_{\Theta \bar{\Omega}} \cc ^{-1} && X_{\Theta \Lambda}
\end{pmatrix}, \\
\tilde{\mathbf{X}}_{\Xi \xi} &= \begin{pmatrix}
X_{\Omega \omega} && X_{\Omega \bar{\omega}}\cc ^{-1} && X_{\Omega \lambda} \\
\cc X_{\bar{\Omega} \omega} && \cc X_{\bar{\Omega} \bar{\omega}} \cc^{-1} && \cc X_{\bar{\Omega} \lambda} \\
X_{\Lambda \omega} && X_{\Lambda \bar{\omega}} \cc ^{-1} && X_{\Lambda \lambda}
\end{pmatrix}, \\
\mathbf{\Delta}_{\Phi} &= \begin{pmatrix}
X_{\Sigma \Sigma} && -P_{\Sigma^{*}}^2+M_\Sigma^2+X_{\Sigma \Sigma ^{*}} &&  X_{\Sigma \Theta} \\
-P_{\Sigma}^2+M_\Sigma^2+X_{\Sigma ^* \Sigma} && X_{\Sigma ^* \Sigma ^*} &&  X_{\Sigma^* \Theta} \\
X_{\Theta \Sigma} && X_{\Theta \Sigma ^*} && -P_\Theta^2+M_\Theta^2+X_{\Theta \Theta}
\end{pmatrix},
\label{eq:DeltaPhi}
\end{align}
with similar definitions for $\Phi\rightarrow \phi$ and
$\Xi\rightarrow \xi$.  Here $P^\mu \equiv i D^\mu$, with $D^\mu$ being
the gauge-covariant derivative, is a matrix diagonal in field space
for which the subscript indicates which gauge group generators are to
be used.  Furthermore we have defined
\begin{align}
(X_{A B})_{ij}\equiv -\frac{\delta ^2 \Lagint}{\delta A_i \delta B_j},
\end{align}
where $\Lagint$ is the interaction Lagrangian of the UV theory and $A$
and $B$ designate arbitrary (scalar or fermionic) fields, if not
stated otherwise.  Here the indices $i$ and $j$ collectively denote all of the indices carried by the fields $A$ and $B$. It shall be noted that if $P^\mu _\Omega$ contains
generators $T^a _r$ of a representation $r$, then
$P^\mu_{\ccfield{\Omega}}$ contains the generators of the conjugate
representation $\bar{r}$, denoted by $T^{a} _{\bar{r}}$. The same
holds for the generators contained in $P^\mu _\Sigma$ and
$P^\mu _{\Sigma^*}$.  Note also that \eqref{eq:variation-part2} is in
principle equivalent to the quadratic term in
\eqref{eq:actionExpansion} with the difference being that in
\eqref{eq:actionExpansion} all scalar fields are assumed to be real,
while in \eqref{eq:variation-part2} complex and real fields are
separate.  The different signs in the fermionic terms in
\eqref{eq:variation-part1} result from using the anti-commutation
relation between fermions and derivatives w.r.t.\ fermions.

Before proceeding it is convenient to define
\begin{align}
\tilde{\mathds{1}}\equiv \begin{pmatrix}
0 && \mathds{1} && 0 \\
\mathds{1} && 0 && 0 \\
0 && 0 && \mathds{1}
\end{pmatrix},
\label{eq:fermID}
\end{align}
and rewrite \eqref{eq:initialDeltaXi} as
\begin{align}
\mathbf{\Delta}_\Xi &= \cc \tilde{\mathds{1}} (\slashed{P}-M_\Xi) +\tilde{\mathbf{X}}_{\Xi \Xi}, 
\end{align}
where
\begin{align}
\slashed{P}-M_\Xi &= \begin{pmatrix}
\slashed{P}_{\Omega}-M_\Omega && 0 && 0 \\
0 && \slashed{P}_{\ccfield{\Omega}}-M_\Omega && 0 \\
0 && 0 && \slashed{P}_\Lambda-M_\Lambda
\end{pmatrix},\\
\tilde{\mathbf{X}}_{\Xi \Xi} &= \begin{pmatrix}
X_{\omega \omega} && X_{\Omega \bar{\Omega}}\cc^{-1} && X_{\Omega \Lambda} \\
\cc X_{\bar{\Omega} \Omega} && \cc X_{\bar{\omega} \bar{\omega}} \cc^{-1} && \cc X_{\bar{\Omega} \Lambda} \\
X_{\Lambda \Omega} && X_{\Lambda \bar{\Omega}}\cc ^{-1} &&  X_{\Lambda \Lambda}
\end{pmatrix}.
\end{align}
We rewrite \eqref{eq:DeltaPhi} in a similar way as
\begin{align}
\mathbf{\Delta}_\Phi=
\tilde{\mathds{1}}(-P^2+M^2_\Phi)+\tilde{\mathbf{X}}_{\Phi \Phi},
\end{align}
with
\begin{align}
-P^2+M^2_\Phi &= \begin{pmatrix}
-P_{\Sigma} ^2 + M^2_\Sigma && 0 && 0 \\
0 && -P_{\Sigma^*} ^2 + M^2_{\Sigma^*} && 0 \\
0 && 0 && -P_{\Theta} ^2 + M^2_{\Theta}
\end{pmatrix}, \\
\tilde{\mathbf{X}}_{\Phi \Phi} &= \begin{pmatrix} X_{\Sigma \Sigma} && X_{\Sigma \Sigma ^{*}} &&  X_{\Sigma \Theta} \\
X_{\Sigma ^* \Sigma} && X_{\Sigma ^* \Sigma ^*} &&  X_{\Sigma^* \Theta} \\
X_{\Theta \Sigma} && X_{\Theta \Sigma ^*} && X_{\Theta \Theta}\end{pmatrix}.
\end{align}
The calculation now proceeds by diagonalizing the quadratic variation
in terms of statistics in order to be able to perform the (Gaussian)
path integral. We first eliminate terms that mix scalar fluctuations
and fluctuations of light fermions $\xi$ by rewriting the second
variation as
\begin{align}
\delta ^2 \Lag_\xi ={}& \frac{1}{2} \delta \xi ^T \tilde{\mathbf{X}}_{\xi \Xi} \delta \Xi +\frac{1}{2}\delta \Xi ^T \tilde{\mathbf{X}}_{\Xi \xi} \delta \xi+\frac{1}{2}\delta \xi ^T \mathbf{\Delta}_\xi \delta \xi -\frac{1}{2} \delta \xi ^T \tilde{\mathbf{X}}_{\xi \Phi} \delta \Phi +\frac{1}{2} \delta \Phi ^T \tilde{\mathbf{X}}_{\Phi \xi} \delta \xi \nonumber \\ & -\frac{1}{2} \delta \xi ^T \tilde{\mathbf{X}}_{\xi \phi} \delta \phi+\frac{1}{2} \delta \phi ^T \tilde{\mathbf{X}}_{\phi \xi} \delta \xi \\ 
={}& \frac{1}{2} \left(\delta \xi^T+\left[\delta \Xi^T \tilde{\mathbf{X}}_{\Xi \xi}+\delta \Phi^T \tilde{\mathbf{X}}_{\Phi \xi}+\delta \phi^T \tilde{\mathbf{X}}_{\phi \xi}\right]\overleftarrow{\mathbf{\Delta}}_\xi^{-1}\right)\mathbf{\Delta}_\xi \nonumber \\ & \times \left(\delta \xi+\mathbf{\Delta}_\xi^{-1}\left[\tilde{\mathbf{X}}_{\xi \Xi} \delta \Xi-\tilde{\mathbf{X}}_{\xi \Phi} \delta \Phi-\tilde{\mathbf{X}}_{\xi \phi} \delta \phi\right]\right) \nonumber \\
& -\frac{1}{2} \left[\delta \Xi^T \tilde{\mathbf{X}}_{\Xi \xi}+\delta \Phi^T \tilde{\mathbf{X}}_{\Phi \xi}+\delta \phi^T \tilde{\mathbf{X}}_{\phi \xi}\right]\mathbf{\Delta}_\xi^{-1}\left[\tilde{\mathbf{X}}_{\xi \Xi} \delta \Xi-\tilde{\mathbf{X}}_{\xi \Phi} \delta \Phi-\tilde{\mathbf{X}}_{\xi \phi} \delta \phi\right].
\end{align}
In the last step we have introduced $\mathbf{\Delta}_\xi ^{-1}$, which
is the matrix-valued Green's function of $\mathbf{\Delta}_\xi$.  The
occurring matrix multiplication also implies an integration, that is
\begin{multline}
  \left(\mathbf{\Delta}_\xi^{-1}\left[\tilde{\mathbf{X}}_{\xi \Xi} \delta \Xi-\tilde{\mathbf{X}}_{\xi \Phi} \delta \Phi-\tilde{\mathbf{X}}_{\xi \phi} \delta \phi\right]\right)(x)\equiv \\
  \int \rd^dy \; \mathbf{\Delta}_\xi^{-1}(x,y)\left[\tilde{\mathbf{X}}_{\xi \Xi}(y) \delta \Xi(y)-\tilde{\mathbf{X}}_{\xi \Phi}(y) \delta \Phi(y)-\tilde{\mathbf{X}}_{\xi \phi}(y) \delta \phi(y)\right].
\end{multline}
Similar to $\mathbf{\Delta}_\xi ^{-1}$ we define
$\overleftarrow{\mathbf{\Delta}}_\xi^{-1}$ in such a way that
\begin{align}
\int \rd^dy \; f(y) \overleftarrow{\mathbf{\Delta}}_\xi ^{-1}(y,x) \overleftarrow{\mathbf{\Delta}} _\xi (x)=f(x),
\end{align}
where
$\overleftarrow{\mathbf{\Delta}} _\xi
(x)=-\overleftarrow{\slashed{P}}-M_{\xi}$.  Next, we shift the light
fermion field as
\begin{align}
  \delta \xi' &= \delta \xi+\mathbf{\Delta}_\xi^{-1}\left[\tilde{\mathbf{X}}_{\xi \Xi} \delta \Xi-\tilde{\mathbf{X}}_{\xi \Phi} \delta \Phi-\tilde{\mathbf{X}}_{\xi \phi} \delta \phi\right],
  \label{eq:xishift} \\
  \delta \xi'^T &= \delta \xi^T+\left[\delta \Xi^T \tilde{\mathbf{X}}_{\Xi \xi}+\delta \Phi^T \tilde{\mathbf{X}}_{\Phi \xi}+\delta \phi^T \tilde{\mathbf{X}}_{\phi \xi}\right]\overleftarrow{\mathbf{\Delta}}_\xi^{-1},
  \label{eq:xishift_T}
\end{align}
under which the path integral measure is invariant.  Since $\xi$ is a
multiplet of Majorana-like spinors, the two shifts \eqref{eq:xishift}
and \eqref{eq:xishift_T} are not independent.  The required relation
between the two shifts is proven in \appref{sec:shifts}.  After the
shifts have been performed we arrive at
\begin{align}
\delta ^2 \Lag_\xi
={}& \frac{1}{2} \delta \xi'^{T} \mathbf{\Delta}_\xi \delta \xi'-\frac{1}{2}\delta \Xi^T \tilde{\mathbf{X}}_{\Xi \xi} \mathbf{\Delta}_\xi^{-1} \tilde{\mathbf{X}}_{\xi \Xi} \delta \Xi+\frac{1}{2}\delta \Xi ^T \tilde{\mathbf{X}}_{\Xi \xi} \mathbf{\Delta}_\xi ^{-1} \tilde{\mathbf{X}}_{\xi \Phi} \delta \Phi+\frac{1}{2}\delta \Xi ^T \tilde{\mathbf{X}}_{\Xi \xi} \mathbf{\Delta}_\xi ^{-1} \tilde{\mathbf{X}}_{\xi \phi} \delta \phi \nonumber \\ 
& +\frac{1}{2} \delta \Phi^T \tilde{\mathbf{X}}_{\Phi \xi} \mathbf{\Delta}_\xi ^{-1} \tilde{\mathbf{X}}_{\xi \Xi} \delta \Xi-\frac{1}{2} \delta \Phi^T \tilde{\mathbf{X}}_{\Phi \xi} \mathbf{\Delta}_\xi ^{-1} \tilde{\mathbf{X}}_{\xi \Phi} \delta \Phi-\frac{1}{2} \delta \Phi^T \tilde{\mathbf{X}}_{\Phi \xi} \mathbf{\Delta}_\xi ^{-1} \tilde{\mathbf{X}}_{\xi \phi} \delta \phi \nonumber \\
& +\frac{1}{2} \delta \phi^T \tilde{\mathbf{X}}_{\phi \xi} \mathbf{\Delta}_\xi ^{-1} \tilde{\mathbf{X}}_{\xi \Xi} \delta \Xi-\frac{1}{2} \delta \phi^T \tilde{\mathbf{X}}_{\phi \xi} \mathbf{\Delta}_\xi ^{-1} \tilde{\mathbf{X}}_{\xi \Phi} \delta \Phi-\frac{1}{2} \delta \phi^T \tilde{\mathbf{X}}_{\phi \xi} \mathbf{\Delta}_\xi ^{-1} \tilde{\mathbf{X}}_{\xi \phi} \delta \phi.
\label{eq:original_2nd_variation}
\end{align}
We proceed by eliminating terms
that mix scalar fluctuations and fluctuations of heavy fermions $\Xi$.
It is convenient to first introduce
\begin{align}
\bar{\mathbf{X}}_{A B}&\equiv \tilde{\mathbf{X}}_{A B}-\tilde{\mathbf{X}}_{A \xi}\mathbf{\Delta}_\xi ^{-1} \tilde{\mathbf{X}}_{\xi B},
\label{eq:quantitiesWithTilde_1} \\
\bar{\mathbf{\Delta}}_A&\equiv \mathbf{\Delta}_A-\tilde{\mathbf{X}}_{A \xi}\mathbf{\Delta}_\xi ^{-1} \tilde{\mathbf{X}}_{\xi A},
\label{eq:quantitiesWithTilde_2}
\end{align}
and write the second variation as
\begin{align}
\delta^2 \Lag= \delta^2 \bar{\Lag}_\text{S}+\frac{1}{2}\delta \Xi ^T \bar{\mathbf{\Delta}}_{\Xi} \delta \Xi- \frac{1}{2} \delta \Xi ^T \bar{\mathbf{X}}_{\Xi \Phi} \delta \Phi+\frac{1}{2} \delta \Phi ^T \bar{\mathbf{X}}_{\Phi \Xi} \delta \Xi -\frac{1}{2} \delta \Xi ^T \bar{\mathbf{X}} _{\Xi \phi} \delta \phi+\frac{1}{2} \delta \phi ^T \bar{\mathbf{X}} _{\phi \Xi} \delta \Xi.
\label{eq:d2_Lag_step_1}
\end{align}
In \eqref{eq:d2_Lag_step_1} the first term on the r.h.s.,
$\delta^2 \bar{\Lag}_\text{S}$, is obtained by replacing
$\tilde{\mathbf{X}}_{A B}$ and $\mathbf{\Delta}_A$ in
$\delta^2 \Lag_\text{S}$ via the relations
\eqref{eq:quantitiesWithTilde_1}--\eqref{eq:quantitiesWithTilde_2}.
By shifting the $\delta \Xi$ in a similar way,
\begin{align}
  \delta \Xi' &= \delta \Xi-\bar{\mathbf{\Delta}}_\Xi^{-1}\left[\bar{\mathbf{X}}_{\Xi \Phi} \delta \Phi+\bar{\mathbf{X}}_{\Xi \phi} \delta \phi\right], \\
  \delta \Xi'^T &= \delta \Xi ^T+\left[\delta \Phi^T \bar{\mathbf{X}}_{\Phi \Xi} +\delta \phi ^T \bar{\mathbf{X}}_{\phi \Xi} \right] \overleftarrow{\bar{\mathbf{\Delta}}}_\Xi^{-1}
\label{eq:Xishift}
\end{align}
one finds
\begin{align}
  \delta^2 \Lag ={}&
  -\frac{1}{2} \delta \Phi ^T (\bar{\mathbf{\Delta}}_{\Phi}-\bar{\mathbf{X}}_{\Phi \Xi}\bar{\mathbf{\Delta}}^{-1}_{\Xi}\bar{\mathbf{X}}_{\Xi \Phi}) \delta \Phi
  -\frac{1}{2} \delta \phi ^T (\bar{\mathbf{\Delta}}_{\phi}-\bar{\mathbf{X}}_{\phi \Xi}\bar{\mathbf{\Delta}}^{-1}_{\Xi}\bar{\mathbf{X}}_{\Xi \phi}) \delta \phi \nonumber \\ 
  & -\frac{1}{2} \delta \Phi ^T (\bar{\mathbf{X}}_{\Phi \phi}-\bar{\mathbf{X}}_{\Phi \Xi}\bar{\mathbf{\Delta}}^{-1}_{\Xi}\bar{\mathbf{X}}_{\Xi \phi}) \delta \phi \nonumber \\ 
  & -\frac{1}{2}\delta \phi ^T (\bar{\mathbf{X}}_{\phi \Phi}-\bar{\mathbf{X}}_{\phi \Xi}\bar{\mathbf{\Delta}}^{-1}_{\Xi}\bar{\mathbf{X}}_{\Xi \Phi}) \delta \Phi+\frac{1}{2} \delta \xi'^{T} \mathbf{\Delta}_\xi \delta \xi'+\frac{1}{2} \delta \Xi'^{T} \bar{\mathbf{\Delta}}_\Xi \delta \Xi' \\
={}& -\frac{1}{2} \begin{pmatrix}
\delta \Phi^T && \delta \phi^T
\end{pmatrix}
\begin{pmatrix}
\bar{\mathbf{\Delta}}_{\Phi}-\bar{\mathbf{X}}_{\Phi \Xi}\bar{\mathbf{\Delta}}^{-1}_{\Xi}\bar{\mathbf{X}}_{\Xi \Phi} && \bar{\mathbf{X}}_{\Phi \phi}-\bar{\mathbf{X}}_{\Phi \Xi}\bar{\mathbf{\Delta}}^{-1}_{\Xi}\bar{\mathbf{X}}_{\Xi \phi} \\
\bar{\mathbf{X}}_{\phi \Phi}-\bar{\mathbf{X}}_{\phi \Xi}\bar{\mathbf{\Delta}}^{-1}_{\Xi}\bar{\mathbf{X}}_{\Xi \Phi} &&  \bar{\mathbf{\Delta}}_{\phi}-\bar{\mathbf{X}}_{\phi \Xi}\bar{\mathbf{\Delta}}^{-1}_{\Xi}\bar{\mathbf{X}}_{\Xi \phi}
\end{pmatrix}
\begin{pmatrix}
\delta \Phi \\
\delta \phi
\end{pmatrix}\nonumber \\ 
& +\frac{1}{2} \delta \xi'^{T} \mathbf{\Delta}_\xi \delta \xi'+\frac{1}{2} \delta \Xi'^{T} \bar{\mathbf{\Delta}}_\Xi \delta \Xi' \\ 
\equiv{}& -\frac{1}{2}\begin{pmatrix}
\delta \Phi^T && \delta \phi^T
\end{pmatrix} \fluctS
 \begin{pmatrix}
\delta \Phi \\
\delta \phi
\end{pmatrix} +\frac{1}{2} \delta \xi'^{T} \mathbf{\Delta}_\xi \delta \xi'+\frac{1}{2} \delta \Xi'^{T} \bar{\mathbf{\Delta}}_\Xi \delta \Xi' \\
\equiv{}& \delta^2\Lag_{\text{SF}} + \delta^2\Lag_{\text{F}}
\label{eq:second_var}
\end{align}
with
\begin{align}
  \delta^2\Lag_{\text{SF}} &= -\frac{1}{2}
  \begin{pmatrix}\delta \Phi^T && \delta \phi^T\end{pmatrix} \fluctS
  \begin{pmatrix}\delta \Phi \\ \delta \phi\end{pmatrix}, \\
  \delta^2\Lag_{\text{F}} &= \frac{1}{2} \delta \xi'^{T} \mathbf{\Delta}_\xi \delta \xi'
  + \frac{1}{2} \delta \Xi'^{T} \bar{\mathbf{\Delta}}_\Xi \delta \Xi'.
\end{align}
At this point there are no terms including both a scalar and a
fermionic fluctuation and the path integrals over scalars and fermions
can be performed separately. As has been pointed out in
\cite{Fuentes-Martin:2016uol} it is convenient to diagonalize the
scalar part such that
\begin{align}
\fluctS = \begin{pmatrix}
\hat{\mathbf{\Delta}}_\Phi-\hat{\mathbf{X}}_{\Phi \phi} \hat{\mathbf{\Delta}}_{\phi}^{-1}\hat{\mathbf{X}}_{\phi \Phi} && 0 \\
0 && \hat{\mathbf{\Delta}}_\phi
\end{pmatrix},
\end{align}
where
\begin{align}
\hat{\mathbf{\Delta}}_A &= \bar{\mathbf{\Delta}}_{A}-\bar{\mathbf{X}}_{A \Xi}\bar{\mathbf{\Delta}}^{-1}_{\Xi}\bar{\mathbf{X}}_{\Xi A},\\
\hat{\mathbf{X}}_{A B} &= \bar{\mathbf{X}}_{A B}-\bar{\mathbf{X}}_{A \Xi}\bar{\mathbf{\Delta}}^{-1}_{\Xi}\bar{\mathbf{X}}_{\Xi B},
\end{align}
with $A,B \in \{\phi, \Phi\}$.  The contribution from this mixed
scalar/fermionic part to the effective action is then given by
\begin{align}
\Lag_{\EFT\text{,SF}}^\text{1\Loop}&=\frac{i}{2}\int \frac{\rd^d q}{(2\pi)^d}\left[ \tr \log 
\left(\hat{\mathbf{\Delta}}_\Phi-\hat{\mathbf{X}}_{\Phi \phi} \hat{\mathbf{\Delta}}_{\phi}^{-1}\hat{\mathbf{X}}_{\phi \Phi}\right) + \tr \log \left.
\hat{\mathbf{\Delta}}_\phi\right]\right \rvert ^{P\rightarrow P-q} _\text{hard}
\label{eq:scalarcontribution}
\end{align}
and it can be calculated using a covariant
derivative expansion as outlined in e.g.\
\cite{Zhang:2016pja}. However, care has to be taken since
$\hat{\mathbf{\Delta}}_\phi$ contains contributions from heavy
fermions and hence does not vanish completely in the hard region of
the momentum integration. The corresponding contributions can be
calculated by using
\begin{align}
\log \det \left(\bar{\mathbf{\Delta}}_{\phi}-\bar{\mathbf{X}}_{\phi \Xi}\bar{\mathbf{\Delta}}^{-1}_{\Xi}\bar{\mathbf{X}}_{\Xi \phi}\right)=\log \det \left(\bar{\mathbf{\Delta}}_{\phi}\right)+\log \det \left(\mathds{1}-\bar{\mathbf{\Delta}}_{\phi}^{-1}\bar{\mathbf{X}}_{\phi \Xi}\bar{\mathbf{\Delta}}^{-1}_{\Xi}\bar{\mathbf{X}}_{\Xi \phi}\right),
\end{align}
where the first term on the right hand side vanishes in the hard
region as it only contains contributions from light fields.

Since a lot of terms are generated when re-expressing the hatted and
barred quantities in terms of the quantities arising in the original
variation \eqref{eq:original_2nd_variation} we abstain from writing
out the result explicitly.  It is, however, useful to consider the
expansion of the hatted operators in order to understand the
ingredients entering the final result. In particular we will show that
it is possible to absorb all explicit factors of $\tilde{\mathds{1}}$
and $\cc$ by appropriate re-definitions of
$\tilde{\mathbf{X}}_{A B}$.  In order to achieve that we first
expand
$(\mathbf{\Delta}_\xi ^{-1})_{P_\mu\to P_\mu-q_\mu}\equiv
\mathbf{\Delta}_\xi ^{-1}(q)$ as
\begin{align}
\mathbf{\Delta}_\xi ^{-1}(q)&=\left[\cc \tilde{\mathds{1}}(\slashed{P}-\slashed{q}-M_\xi) +\tilde{\mathbf{X}}_{\xi \xi}\right]^{-1} \\
&= \left[\mathds{1}-\left(-\slashed{q}-M_{\xi}\right)^{-1}\tilde{\mathds{1}}\cc^{-1}\left(-\cc \tilde{\mathds{1}}\slashed{P}-\tilde{\mathbf{X}}_{\xi \xi}\right)\right]\left(-\slashed{q}-M_{\xi}\right)^{-1}\tilde{\mathds{1}}\cc^{-1} \\
&= \sum_{n=0} ^\infty \left[\left(-\slashed{q}-M_{\xi}\right)^{-1}\tilde{\mathds{1}}\cc^{-1}\left(-\cc \tilde{\mathds{1}}\slashed{P}-\tilde{\mathbf{X}}_{\xi \xi}\right) \right]^n \left(-\slashed{q}-M_{\xi}\right)^{-1}\tilde{\mathds{1}}\cc^{-1} \\
&= \sum_{n=0} ^\infty \left[\left(-\slashed{q}-M_{\xi}\right)^{-1}\left(-\slashed{P}-\mathbf{X}_{\xi \xi}\right) \right]^n \left(-\slashed{q}-M_{\xi}\right)^{-1}\tilde{\mathds{1}}\cc^{-1},
\label{eq:DeltaxiInv}
\end{align}
where we defined 
\begin{align}
\mathbf{X}_{\xi \xi}\equiv \tilde{\mathds{1}}\cc ^{-1} \tilde{\mathbf{X}}_{\xi \xi}.
\end{align}
Then
\eqref{eq:quantitiesWithTilde_1}--\eqref{eq:quantitiesWithTilde_2}
become
\begin{align}
\bar{\mathbf{X}}_{A B}&= \tilde{\mathbf{X}}_{A B}-\tilde{\mathbf{X}}_{A \xi}\sum _{n=0} ^{\infty} \left[\left(-\slashed{q}-M_{\xi}\right) ^{-1} \left(-\mathbf{X} _{\xi \xi}-\slashed{P}\right) \right]^n \left(-\slashed{q}-M_{\xi}\right) ^{-1} \mathbf{X}_{\xi B}, \\
\bar{\mathbf{\Delta}}_A&= \mathbf{\Delta}_A-\tilde{\mathbf{X}}_{A \xi}\sum _{n=0} ^{\infty} \left[\left(-\slashed{q}-M_{\xi}\right) ^{-1} \left(-\mathbf{X} _{\xi \xi}-\slashed{P}\right) \right]^n \left(-\slashed{q}-M_{\xi}\right) ^{-1} \mathbf{X}_{\xi A},
\end{align}
where we introduced
$\mathbf{X}_{\xi B}\equiv \cc ^{-1}\tilde{\mathds{1}}
\tilde{\mathbf{X}}_{\xi B}$. Next we consider
\begin{align}
\bar{\mathbf{\Delta}}_\Xi ^{-1}(q) &= \Bigg[\cc \tilde{\mathds{1}}\left(-\slashed{q}-M_{\Xi}\right) +\cc \tilde{\mathds{1}} \slashed{P}+\tilde{\mathbf{X}}_{\Xi \Xi} \nonumber \\
 &~~~~~~~ -\tilde{\mathbf{X}}_{\Xi \xi}\sum _{n=0} ^{\infty} \left[\left(-\slashed{q}-M_{\xi}\right) ^{-1} \left(-\mathbf{X} _{\xi \xi}-\slashed{P}\right) \right]^n \left(-\slashed{q}-M_{\xi}\right) ^{-1} \mathbf{X}_{\xi \Xi} \Bigg]^{-1} \\
&=\sum _{m=0} ^{\infty} \left\{\mathcal{K}_\Xi ^{-1}  \left(-\mathbf{X} _{\Xi \Xi}-\slashed{P}\right) + \mathcal{K}_\Xi ^{-1}  \mathbf{X}_{\Xi \xi}\sum _{n=0} ^{\infty} \left[\mathcal{K}_\xi ^{-1} \left(-\mathbf{X} _{\xi \xi}-\slashed{P}\right) \right]^n \mathcal{K}_\xi ^{-1} \mathbf{X}_{\xi \Xi}  \right\}^m \mathcal{K}_\Xi ^{-1} \cc ^{-1} \tilde{\mathds{1}},
\label{eq:DeltaXiTildeInv}
\end{align}
where
\begin{align}
  \mathcal{K}_A &\equiv \left(-\slashed{q}-M_A\right), \\
  \mathbf{X}_{\Xi \xi} &\equiv \cc ^{-1} \tilde{\mathds{1}} \tilde{\mathbf{X}}_{\Xi \xi}.
\end{align}
Note that in \eqref{eq:DeltaxiInv} and \eqref{eq:DeltaXiTildeInv} the
expressions for $\mathbf{\Delta}_\xi ^{-1}$ and
$\bar{\mathbf{\Delta}}^{-1}_{\Xi}$ contain the factor
$\cc ^{-1} \tilde{\mathds{1}}$ on the very right.  This means that in
the combination
\begin{align}
\bar{\mathbf{\Delta}}^{-1}_{\Xi}\bar{\mathbf{X}}_{\Xi B}&=\bar{\mathbf{\Delta}}^{-1}_{\Xi} (\tilde{\mathbf{X}}_{\Xi B}-\tilde{\mathbf{X}}_{\Xi \xi}\mathbf{\Delta}_\xi ^{-1} \tilde{\mathds{1}} \cc \cc^{-1} \tilde{\mathds{1}} \tilde{\mathbf{X}} _{\xi B}) \\
&=\bar{\mathbf{\Delta}}^{-1}_{\Xi} \tilde{\mathds{1}} \cc \cc^{-1} \tilde{\mathds{1}} (\tilde{\mathbf{X}}_{\Xi B}-\tilde{\mathbf{X}}_{\Xi \xi}\mathbf{\Delta}_\xi ^{-1} \tilde{\mathds{1}} \cc \cc^{-1} \tilde{\mathds{1}} \tilde{\mathbf{X}} _{\xi B}) \\
&=\bar{\mathbf{\Delta}}^{-1}_{\Xi} \tilde{\mathds{1}} \cc (\mathbf{X}_{\Xi B}-\mathbf{X}_{\Xi \xi}\mathbf{\Delta}_\xi ^{-1} \tilde{\mathds{1}} \cc \mathbf{X}_{\xi B}),
\end{align}
all appearances of $\cc$ and $\tilde{\mathds{1}}$ cancel once
$\bar{\mathbf{\Delta}}^{-1}_{\Xi}$ and $\mathbf{\Delta}_\xi ^{-1}$ are
inserted and $\tilde{\mathbf{X}}_{A B}$ is expressed in terms of
$\mathbf{X}_{A B}$ with
$\mathbf{X}_{A B}=\cc^{-1} \tilde{\mathds{1}}
\tilde{\mathbf{X}}_{A B}$.  A similar property holds for
$\tilde{\mathbf{X}}_{\Phi B}$ and $\tilde{\mathbf{X}}_{\phi B}$,
which only appear as
$\mathbf{X}_{\Phi B}=\tilde{\mathds{1}}\tilde{\mathbf{X}}_{\Phi
  B}$ and
$\mathbf{X}_{\phi B}=\tilde{\mathds{1}}\tilde{\mathbf{X}}_{\phi
  B}$.
% That this is the case follows from the fact that both
% $\mathbf{\Delta} _{\phi}$ and $\mathbf{\Delta} _{\Phi}$ contain a
% $\tilde{\mathds{1}}$, but no $\cc$.
Hence, the result can be expressed
entirely through the matrices $\mathbf{X}_{A B}$ and neither
$\tilde{\mathds{1}}$ nor $\cc$ explicitly appears in the final
operator structures.

To complete the calculation we need to compute the purely fermionic
part of the second variation \eqref{eq:second_var}, which reads
\begin{align}
  \delta^2 \Lag_\text{F} = \frac{1}{2}\delta \Xi'^T \bar{\mathbf{\Delta}}_\Xi \delta \Xi'
  + \frac{1}{2} \delta \xi'^T \mathbf{\Delta}_\xi \delta \xi'.
\end{align}
Again, we are only interested in the contribution from the hard region
where the light only part $\mathbf{\Delta} _\xi$ does not
contribute. Hence we only need to consider
$\bar{\mathbf{\Delta}}_\Xi$. We find
\begin{align}
  \tr \log \Big(&\mathbf{\Delta}_\Xi(q) - \mathbf{X}_{\Xi \xi}\Delta_\xi ^{-1}(q) \mathbf{X}_{\xi \Xi}\Big)\nonumber \\ &= \tr \log \left(\cc \tilde{\mathds{1}} \mathcal{K}_\Xi+\cc \tilde{\mathds{1}} \slashed{P}+\tilde{\mathbf{X}}_{\Xi \Xi}-\mathbf{X}_{\Xi \xi}\mathbf{\Delta}_\xi ^{-1}(q) \tilde{\mathbf{X}}_{\xi \Xi}\right) \\
  &= \tr \log \left( \cc \tilde{\mathds{1}}\mathcal{K}_\Xi \right) + \tr \log \left[ \mathds{1}-\mathcal{K}_\Xi^{-1}  \left(-\slashed{P}-\mathbf{X}_{\Xi \Xi}+\mathbf{X}_{\Xi \xi}\mathbf{\Delta}_\xi ^{-1}(q) \tilde{\mathbf{X}}_{\xi \Xi} \right) \right],
\label{eq:trlog_fermionic}
\end{align}
where the first term on the r.h.s.\ of \eqref{eq:trlog_fermionic} is
absorbed in the normalization of the path integral. Inserting
$\mathbf{\Delta}_\xi ^{-1}(q)$ from \eqref{eq:DeltaxiInv} yields
\begin{align}
\Lag_\text{\EFT,F}^{1\Loop} &= \frac{i}{2} \sum_{n=1}^{\infty} \frac{1}{n} \tr \left[\mathcal{K}_\Xi^{-1}\left(-\slashed{P}-\mathbf{X}_{\Xi \Xi}+\mathbf{X} _{\Xi \xi}\sum _{m=0} ^{\infty} \left[\mathcal{K} _\xi ^{-1} \mathbf{X} _{\xi \xi} \right]^m \mathcal{K} _\xi ^{-1} \mathbf{X} _{\xi \Xi} \right) \right]^n.
\end{align}
In order to obtain the final UOLEA from the sum
\begin{align}
  \Lag_{\EFT}^{1\Loop} = \Lag_{\EFT\text{,SF}}^{1\Loop} + \Lag_\text{\EFT,F}^{1\Loop}
  \label{eq:UOLEA_final}
\end{align}
one needs to expand all functional traces on the r.h.s.\ of
\eqref{eq:UOLEA_final} to a given mass dimension and calculate the
coefficients and operator structures.  In this expansion we keep
$P^\mu$ as a whole to obtain a manifestly gauge-invariant effective
Lagrangian.
It can be shown, by using the Baker-Campbell-Hausdorff
formula, that every $P_\mu$ appears in commutators of the form
$[P_\mu,\bullet]$ \cite{Gaillard:1985uh,Cheyette:1987qz}. To combine
all $P^\mu$ operators into commutators one can either explicitly use
the Baker-Campbell-Hausdorff formula in the calculation as was done in
\cite{Drozd:2015rsp} or construct a basis for these commutators and
then solve a system of equations to fix the coefficients of the basis
elements as was pointed out in \cite{Zhang:2016pja}.  In this
publication the second method was deployed.
Our final expression for $\Lag_{\EFT}^{1\Loop}$ is contained in the
ancillary file \texttt{UOLEA.m} in the arXiv submission of this
publication and will be described further in the next section.

\section{Discussion of the result}
\label{sec:results}

\subsection{Published operators and coefficients}
\label{sec:results_ops_coeffs}

In the following we describe the calculated scalar/fermionic
operators, which we publish in the ancillary file \texttt{UOLEA.m} in
the arXiv submission of this publication.  The file contains the
following four lists:
\begin{itemize}
\item \texttt{mixedLoopsNoP}: Mixed scalar/fermionic operators without
  $P^\mu$.
\item \texttt{mixedLoopsWithP}: Mixed scalar/fermionic operators with
  $P^\mu$.
\item \texttt{fermionicLoopsNoP}: Purely fermionic operators without
  $P^\mu$.
\item \texttt{fermionicLoopsWithP}: Purely fermionic operators with
  $P^\mu$.
\end{itemize}
For convenience, the additional list \texttt{uolea} is defined, which
is the union of the four lists from above.  The lists contain the
calculated operators in the form
$\{F^\alpha(M_i,M_j,\dots),\mathcal{O}^\alpha_{ij\cdots}\}$, where $F^\alpha(M_i,M_j,\dots)$
is the coefficient of the operator $\mathcal{O}^\alpha_{ij\cdots}$, which is
expressed through the integrals
$\ZI [q^{2n_c}]^{n_i n_j \dots n_L} _{i j \dots 0}$ defined in
\appref{sec:loop_functions}. The operators $\mathcal{O}^\alpha_{ij\cdots}$ are
expressed in terms of the symbols
$X[\text{A},\text{B}][i,j]$, with
$\text{A}, \text{B}\in \{\text{S},\text{s},\text{F},\text{f}\}$, which
correspond to the matrices defined in \secref{sec:calc} as follows:
\begin{align*}
X[\text{S},\text{F}] &\equiv \mathbf{X}_{\Phi \Xi}= \begin{pmatrix}
X_{\Sigma ^* \Omega} && X_{\Sigma ^* \bar{\Omega}} \cc ^{-1} && X_{\Sigma ^* \Lambda} \\
X_{\Sigma \Omega} && X_{\Sigma \bar{\Omega}} \cc ^{-1} && X_{\Sigma \Lambda} \\
X_{\Theta \Omega} && X_{\Theta \bar{\Omega}} \cc ^{-1} && X_{\Theta \Lambda}
\end{pmatrix}, \nonumber \\
X[\text{s},\text{F}] &\equiv \mathbf{X}_{\phi \Xi}= \begin{pmatrix}
X_{\sigma ^* \Omega} && X_{\sigma ^* \bar{\Omega}} \cc ^{-1} && X_{\sigma ^* \Lambda} \\
X_{\sigma \Omega} && X_{\sigma \bar{\Omega}} \cc ^{-1} && X_{\sigma \Lambda} \\
X_{\theta \Omega} && X_{\theta \bar{\Omega}} \cc ^{-1} && X_{\theta \Lambda}
\end{pmatrix}, \nonumber \\
X[\text{S},\text{f}] &\equiv \mathbf{X}_{\Phi \xi}=\begin{pmatrix}
X_{\Sigma ^* \omega} && X_{\Sigma ^* \bar{\omega}} \cc ^{-1} && X_{\Sigma ^* \lambda} \\
X_{\Sigma \omega} && X_{\Sigma \bar{\omega}} \cc ^{-1} && X_{\Sigma \lambda} \\
X_{\Theta \omega} && X_{\Theta \bar{\omega}} \cc ^{-1} && X_{\Theta \lambda}
\end{pmatrix}, \nonumber \\
X[\text{s},\text{f}] &\equiv \mathbf{X}_{\phi \xi}= \begin{pmatrix}
X_{\sigma ^* \omega} && X_{\sigma ^* \bar{\omega}} \cc ^{-1} && X_{\sigma ^* \lambda} \\
X_{\sigma \omega} && X_{\sigma \bar{\omega}} \cc ^{-1} && X_{\sigma \lambda} \\
X_{\theta \omega} && X_{\theta \bar{\omega}} \cc ^{-1} && X_{\theta \lambda}
\end{pmatrix}, \nonumber \\
X[\text{F},\text{S}] &\equiv \mathbf{X}_{\Xi \Phi} = \begin{pmatrix}
 X_{\bar{\Omega} \Sigma} &&  X_{\bar{\Omega} \Sigma ^*} &&  X_{\bar{\Omega} \Theta} \\ 
\cc^{-1} X_{\Omega \Sigma} && \cc^{-1} X_{\Omega \Sigma^{*}} &&  \cc^{-1} X_{\Omega \Theta} \\ 
\cc^{-1} X_{\Lambda \Sigma} && \cc^{-1} X_{\Lambda \Sigma ^*} && \cc^{-1} X_{\Lambda \Theta}
\end{pmatrix}, \nonumber \\
X[\text{f},\text{S}] &\equiv \mathbf{X}_{\xi \Phi} = \begin{pmatrix}
 X_{\bar{\omega} \Sigma} &&  X_{\bar{\omega} \Sigma ^*} &&  X_{\bar{\omega} \Theta} \\ 
\cc^{-1} X_{\omega \Sigma} && \cc^{-1} X_{\omega \Sigma^{*}} &&  \cc^{-1} X_{\omega \Theta} \\ 
\cc^{-1} X_{\lambda \Sigma} && \cc^{-1} X_{\lambda \Sigma ^*} && \cc^{-1} X_{\lambda \Theta}
\end{pmatrix}, \nonumber \\
X[\text{F},\text{s}] &\equiv \mathbf{X}_{\Xi \phi} = \begin{pmatrix}
 X_{\bar{\Omega} \sigma} &&  X_{\bar{\Omega} \sigma ^*} &&  X_{\bar{\Omega} \theta} \\ 
\cc^{-1} X_{\Omega \sigma} && \cc^{-1} X_{\Omega \sigma^{*}} &&  \cc^{-1} X_{\Omega \theta} \\ 
\cc^{-1} X_{\Lambda \sigma} && \cc^{-1} X_{\Lambda \sigma ^*} && \cc^{-1} X_{\Lambda \theta}
\end{pmatrix}, \nonumber \\
X[\text{f},\text{s}] &\equiv \mathbf{X}_{\xi \phi} = \begin{pmatrix}
 X_{\bar{\omega} \sigma} &&  X_{\bar{\omega} \sigma ^*} &&  X_{\bar{\omega} \theta} \\ 
\cc^{-1} X_{\omega \sigma} && \cc^{-1} X_{\omega \sigma^{*}} &&  \cc^{-1} X_{\omega \theta} \\ 
\cc^{-1} X_{\lambda \sigma} && \cc^{-1} X_{\lambda \sigma ^*} && \cc^{-1} X_{\lambda \theta}
\end{pmatrix}, \nonumber \\
X[\text{F},\text{F}] &\equiv \mathbf{X}_{\Xi \Xi}=\begin{pmatrix}
X_{\bar{\Omega} \Omega} && X_{\bar{\Omega} \bar{\Omega}} \cc^{-1} &&  X_{\bar{\Omega} \Lambda} \\
\cc^{-1} X_{\Omega \Omega} &&  \cc ^{-1} X_{\Omega \bar{\Omega}}\cc^{-1} && \cc ^{-1} X_{\Omega \Lambda} \\
\cc ^{-1} X_{\Lambda \Omega} && \cc ^{-1} X_{\Lambda \bar{\Omega}}\cc ^{-1} &&  \cc ^{-1} X_{\Lambda \Lambda}
\end{pmatrix}, \nonumber \\
X[\text{f},\text{f}] &\equiv \mathbf{X}_{\xi \xi}=\begin{pmatrix}
X_{\bar{\omega} \omega} && X_{\bar{\omega} \bar{\omega}} \cc ^{-1} &&  X_{\bar{\omega} \lambda} \\
\cc ^{-1} X_{\omega \omega} &&  \cc ^{-1} X_{\omega \bar{\omega}}\cc^{-1} && \cc ^{-1} X_{\omega \lambda} \\
\cc ^{-1} X_{\lambda \omega} && \cc ^{-1} X_{\lambda \bar{\omega}}\cc ^{-1} &&  \cc ^{-1} X_{\lambda \lambda}
\end{pmatrix}, \nonumber \\
X[\text{F},\text{f}] &\equiv \mathbf{X}_{\Xi \xi}=\begin{pmatrix}
X_{\bar{\Omega} \omega} && X_{\bar{\Omega} \bar{\omega}} \cc^{-1} &&  X_{\bar{\Omega} \lambda} \\
\cc ^{-1} X_{\Omega \omega} && \cc ^{-1} X_{\Omega \bar{\omega}}\cc ^{-1} && \cc^{-1} X_{\Omega \lambda} \\
\cc ^{-1} X_{\Lambda \omega} && \cc ^{-1} X_{\Lambda \bar{\omega}} \cc ^{-1} &&  \cc^{-1} X_{\Lambda \lambda}
\end{pmatrix}, \nonumber \\
X[\text{f},\text{F}] &\equiv \mathbf{X}_{\xi \Xi}=\begin{pmatrix}
X_{\bar{\omega} \Omega} && X_{\bar{\omega} \bar{\Omega}} \cc ^{-1} &&  X_{\bar{\omega} \Lambda} \\
\cc ^{-1} X_{\omega \Omega} && \cc ^{-1} X_{\omega \bar{\Omega}}\cc ^{-1} && \cc^{-1} X_{\omega \Lambda} \\
\cc ^{-1} X_{\lambda \Omega} && \cc ^{-1} X_{\lambda \bar{\Omega}} \cc ^{-1} &&  \cc^{-1} X_{\lambda \Lambda}
\end{pmatrix}, \nonumber \\
X[\text{S},\text{S}] &\equiv \mathbf{X}_{\Phi \Phi}=\begin{pmatrix} 
X_{\Sigma ^* \Sigma} && X_{\Sigma ^* \Sigma ^*} &&  X_{\Sigma^* \Theta} \\
X_{\Sigma \Sigma} && X_{\Sigma \Sigma ^{*}} &&  X_{\Sigma \Theta} \\
X_{\Theta \Sigma} && X_{\Theta \Sigma ^*} && X_{\Theta \Theta}\end{pmatrix}, \nonumber \\
X[\text{S},\text{s}] &\equiv \mathbf{X}_{\Phi \phi}=\begin{pmatrix} 
X_{\Sigma ^* \sigma} && X_{\Sigma ^* \sigma ^*} &&  X_{\Sigma^* \theta} \\
X_{\Sigma \sigma} && X_{\Sigma \sigma ^{*}} &&  X_{\Sigma \theta} \\
X_{\Theta \sigma} && X_{\Theta \sigma ^*} && X_{\Theta \theta}\end{pmatrix}, \nonumber \\
X[\text{s},\text{S}] &\equiv \mathbf{X}_{\phi \Phi}=\begin{pmatrix} 
X_{\sigma ^* \Sigma} && X_{\sigma ^* \Sigma ^*} &&  X_{\sigma^* \Theta} \\
X_{\sigma \Sigma} && X_{\sigma \Sigma ^{*}} &&  X_{\sigma \Theta} \\
X_{\theta \Sigma} && X_{\theta \Sigma ^*} && X_{\theta \Theta}\end{pmatrix}, \nonumber \\
X[\text{s},\text{s}] &\equiv \mathbf{X}_{\phi \phi}=\begin{pmatrix} 
X_{\sigma ^* \sigma} && X_{\sigma ^* \sigma ^*} &&  X_{\sigma^* \theta} \\
X_{\sigma \sigma} && X_{\sigma \sigma ^{*}} &&  X_{\sigma \theta} \\
X_{\theta \sigma} && X_{\theta \sigma ^*} && X_{\theta \theta}\end{pmatrix} .
\end{align*}
The indices $i,j\in\mathbb{N}$ label a specific element of the
respective matrix. The full one-loop effective action is then obtained
as
\begin{align}
\Lag_{\EFT}^{1\Loop} = \kappa\sum_\alpha \sum _{ij \cdots} F^\alpha(M_i,M_j,\dots) \mathcal{O}^\alpha_{ij\cdots},
\label{eq:L_all_generic}
\end{align}
where $\kappa=1/(4\pi)^2$ and the sum over $\alpha$ runs over all operators and their
corresponding coefficients.
Several comments regarding the use of the operators of
\eqref{eq:L_all_generic} are in order. First, no assumptions have
been made about the dependence of the second derivatives
$X_{A B}$ regarding gamma matrices. The result is valid for any
spin $1/2$ spinor structure appearing in these derivatives.
Second, care has to be taken to retain the poles of the coefficients
since the gamma algebra has to be performed in $d = 4 - \eps$
dimensions, which may generate finite contributions when combined with
the poles.  The function \texttt{ExpandEps}, contained in
the ancillary Mathematica file \texttt{LoopFunctions.m} in the arXiv
submission of this paper, can be used to extract these finite
contributions.
Third, some of the coefficients diverge in the case of degenerate
masses if the degenerate limit is not taken carefully. The most
convenient way to deal with degenerate masses may be to first set the
masses equal, which modifies the integrals appearing in the
coefficients $F^\alpha(M_i,M_j,\dots)$, and to then calculate these
integrals using the reduction algorithm implemented in the ancillary 
Mathematica file \texttt{LoopFunctions.m}.
Last, there are no $c_s$ or $c_F$ factors appearing in the final
result, in contrast to \cite{Drozd:2015rsp,Ellis:2017jns,Summ:2018oko}.  In our
formulation these prefactors have been fixed by our treatment of the
different kinds of fields and are absorbed in the coefficients.

\subsection{Infrared and ultra-violet divergences}

It appears that the operator coefficients have infrared divergences,
which might be surprising as the
infrared physics should cancel in the matching. The reason for the
appearance of such poles is the fact that expansion by regions was
used to perform the calculation as discussed in \secref{sec:
  intro}. For a heavy-light loop this means that the one-loop integral
$I_\text{full}$ in the full integration region is split into a part
$I_\text{soft}$, calculated in the soft region, and a part $I_\text{hard}$,
calculated in the hard region,
\begin{align}
  I_\text{full} = I_\text{soft} + I_\text{hard}.
\end{align}
Only the hard part remains, since the soft part is canceled in the
matching by the EFT contribution. For the example of $I_\text{full}$
being finite, a UV-divergence in the soft part of the integration region
cancels with an IR-divergence in the hard part with the condition
\begin{align}
\frac{1}{\epsUV}=\frac{1}{\epsIR},
\label{eq: epsrel}
\end{align}
which assures that scaleless integrals vanish in dimensional
regularization. Since the soft part is removed in the matching, the
IR-divergence of the hard part remains. However, such an IR-divergence
should be interpreted as a subtracted UV-divergence coming from the
EFT as indicated by \eqref{eq: epsrel}. It is not surprising that
these divergences do not cancel in the matching since the UV behavior
of the EFT is modified as compared to the UV-theory. However, since
these genuine UV-divergences may still combine with an $\eps$ from
the gamma algebra to yield finite contributions they must be treated
in the same way as $1/\eps$ poles stemming from the UV behavior of
the UV-theory. After performing the trace and the gamma algebra,
remaining terms containing $1/\eps$ poles can be discarded, which
amounts to performing a matching calculation in the $\MSbar$ scheme.

\subsection{Application to models with massive vector fields}
\label{sec:results_vectors}

The operators calculated in
this paper can be used to treat massive vector fields in Feynman gauge
as described in \cite{Zhang:2016pja}. Furthermore, couplings of
fermions to massless gauge bosons can be correctly accounted for as
well using the same technique and the treatment is complete when the
UV-theory is renormalizable. This follows from the fact that the
gauge-kinetic term of a fermion $\psi$ is linear in the covariant
derivative so that $X_{A_\mu \psi}$ is independent of $P_\mu$. This is
not the case for scalar fields, since the kinetic term is quadratic in
$P_\mu$, which means that even for a renormalizable UV-theory there
are further operators stemming from the coupling of scalar fields to
massless gauge bosons. Of course, once one considers the matching of a
UV-theory that already contains higher dimensional operators with
covariant derivatives to an EFT, further operators arise also for
fermions.  These missing operators all stem from open covariant
derivatives and are currently unknown.

\subsection{Extraction of $\beta$-functions}

As was pointed out in \cite{Henning:2016lyp} functional methods can be
used to calculate $\beta$-functions since they allow for the
computation of the loop-corrected generator of 1PI Green's
functions. To one-loop we have
\begin{align}
  \Gamma[\Phi]=\Gamma^\tree[\Phi]+\Gamma^{1\Loop}[\Phi],
\end{align}
where $\Gamma^\tree[\Phi]=S[\Phi]$ is the tree-level generator of 1PI
Green's functions, which is simply the classical action. Assume that
$\Gamma^\tree[\Phi]$ contains a kinetic term $\mathcal{O}_K[\Phi]$ and
an interaction term $g \mathcal{O}_g[\Phi]$. Then, in general, the
one-loop contribution will contain corrections to these, which depend
on the renormalization scale $\mu$, so that
\begin{align}
  \Gamma[\Phi] \supset \int \rd^4 x \; \big\{a_K(\mu) \mathcal{O}_K[\Phi]+a_g(\mu) \mathcal{O}_g[\Phi]\big\}.
\end{align}
Canonically normalizing the kinetic term
for the field $\Phi$ yields
\begin{align}
  \Gamma[\Phi] \supset \int \rd^4 x \; \big\{\mathcal{O}_K[\Phi]+a'_g(\mu) \mathcal{O}_g[\Phi]\big\},
\end{align}
where
\begin{align}
  \mu \frac{\rd}{\rd\mu}a'_g(\mu)=0
  \label{eq: running equation}
\end{align}
due to the Callan-Symanzik equation
\cite{Callan:1970yg,Symanzik:1970rt}. Eq.~\eqref{eq: running equation}
can be solved for the one-loop $\beta$-function of the coupling $g$.

In a specific sense, the UOLEA represents an expression for
$\Gamma^{1\Loop}$ of a model with operators up to dimension
6, and it can thus be used to calculate the one-loop $\beta$-functions
of all dimension 6 operators for any given Lagrangian as described
above.  In order to calculate $\Gamma^{1\Loop}$, the UOLEA operators
\eqref{eq:L_all_generic} must be re-interpreted as follows: Since one
is interested in the full $\Gamma^{1\Loop}$, a distinction between
heavy and light fields must not be made and all fields shall be treated
as ``heavy'' fields.  As a consequence, the one-loop effective action
of a scalar theory is given by
\begin{align}
  \Gamma[\Phi] = S[\Phi]
  + \frac{i}{2} \log\det\left(-\frac{\delta^2 \Lag_\text{int}}{\delta\Phi\delta\Phi}\right),
  \label{eq:gamma_1L_heavy}
\end{align}
where $\Phi$ represents the collection of all scalar fields contained
in the model.  The expression on the r.h.s.\ of
\eqref{eq:gamma_1L_heavy} can be expanded as outlined e.g.\ in
\cite{Drozd:2015rsp,Henning:2016lyp,Fuentes-Martin:2016uol} and one
arrives at the heavy-only part of the UOLEA \eqref{eq:L_all_generic},
which contains only operators built out of derivatives of the
Lagrangian with respect to ``heavy'' $\Phi$ fields.
This procedure is not restricted to a theory with only
scalars and can also be applied to models with both scalars and
fermions using the heavy-only part of \eqref{eq:L_all_generic}.
However, higher-dimensional operators with covariant derivatives have
not been treated in this work and hence their influence on the running
of the couplings cannot be determined using our result.

\section{Applications}
\label{sec:applications}

\subsection{Integrating out the top quark from the Standard Model}

As a simple first example we consider the corrections to the Higgs
tadpole and mass parameter that arise when integrating out the top
quark from the Standard Model. The considered interaction Lagrangian
shall contain only one coupling
\begin{align}
  \Lag_\SM \supset -\frac{g_t}{\sqrt{2}}h \bar{t}t,
\end{align}
where $h$ denotes the physical Higgs field, $t$ is the top quark and
$g_t$ is the top Yukawa coupling.  The relevant operators of the UOLEA
\eqref{eq:UOLEA_final} are given by
\begin{align}
\frac{1}{\kappa} \Lag_\EFT^{1\Loop} = \tr \Bigg\lbrace & \frac{1}{4} m_{\Xi i} m_{\Xi j}^3 \ZI ^{13} _{ij} [P_\mu,(\mathbf{X}_{\Xi \Xi})_{ij}][P^\mu,(\mathbf{X}_{\Xi \Xi})_{ji}]
\nonumber \\ &  -\frac{1}{2} \ZI[q^4] ^{22} _{ij} \gamma^\nu [P_\mu,(\mathbf{X}_{\Xi \Xi})_{ij}]\gamma_\nu[P^\mu,(\mathbf{X}_{\Xi \Xi})_{ji}]
\nonumber \\ &  - \ZI[q^4] ^{22} _{ij} \gamma^\nu [P_\nu,(\mathbf{X}_{\Xi \Xi})_{ij}]\gamma_\mu[P^\mu,(\mathbf{X}_{\Xi \Xi})_{ji}]
\nonumber \\ & +\frac{1}{2} m_{\Xi i} \ZI ^1 _i (\mathbf{X}_{\Xi \Xi})_{ii}
\nonumber \\ & -\frac{1}{4} m_{\Xi i} m_{\Xi j} \ZI ^{11} _{ij} (\mathbf{X}_{\Xi \Xi})_{ij} (\mathbf{X}_{\Xi \Xi})_{ji}
\nonumber \\ & -\frac{1}{4} \ZI[q^2] ^{11} _{ij} \gamma ^\mu (\mathbf{X}_{\Xi \Xi})_{ij} \gamma_\mu (\mathbf{X}_{\Xi \Xi})_{ji}
\Bigg\rbrace,
\label{eq:UOLEALAG-topout}
\end{align}
where $m_{\Xi i}$ denotes the mass of the $i$th component of $\Xi$. The matrix $(\mathbf{X}_{\Xi \Xi})$ is given by
\begin{align}
(\mathbf{X}_{\Xi \Xi})_{\alpha \beta ij} = \begin{pmatrix}
(X_{\bar{t}t})_{\alpha \beta ij} & 0 \\
0 & \cc^{-1} _{\alpha \rho} (X_{t\bar{t}})_{\rho \sigma ij} \cc^{-1} _{\sigma \beta} 
\end{pmatrix}
= 
-\frac{g_t}{\sqrt{2}}h \delta_{\alpha \beta} \delta_{ij} \mathbf{1}_{2\times 2},
\label{eq:top-derivative}
\end{align}
with $\alpha,\beta = 1,\ldots,4$ being spinor indices and
$i,j = 1,2,3$ being color indices. In \eqref{eq:UOLEALAG-topout} we
included terms with two covariant derivatives in order to obtain the
field-redefinition of the Higgs field that is necessary to canonically
normalize the corresponding Higgs field $\hat{h}$ in the effective
theory.  Since this redefinition arises from the correction to the
kinetic term only, we can set $P^\mu = i\partial ^\mu$.  Inserting
\eqref{eq:top-derivative} into \eqref{eq:UOLEALAG-topout} and
calculating the trace yields
\begin{align}
  \frac{1}{\kappa}\Lag_\EFT ^{1\Loop}  ={}& -3g_t^2 \left(m_t^4 \ZI ^4 _t-2d\ZI [q^4]^4 _{t}-4 \ZI [q^4] ^4 _{t} \right) (\partial_\mu h) (\partial^\mu h) \nonumber \\
  & -3g_t^2 \left(\ZI ^2_t m_t^2+d \ZI [q^2]^2_t \right)h^2-\frac{12}{\sqrt{2}} g_t m_t \ZI^1_t h,
\label{eq: HiggsEFT}
\end{align}
where $d = 4 - \eps = g^\mu _\mu$ has to be retained since the
integrals contain poles in $1/\epsilon$.  The loop functions $\ZI$ are
defined in \appref{sec:loop_functions}. It is customary to introduce
the canonically normalized field $\hat{h}$ which is related to $h$
through
\begin{align}
  \hat{h}=\left(1+\frac{1}{2}\delta Z_h \right) h.
\end{align}
From \eqref{eq: HiggsEFT} one can read off $\delta Z_h$ to be
\begin{align}
  \delta Z_h = -6g_t^2\left(m_t^4 \ZI ^4 _t-2d\ZI [q^4]^4 _{t}-4 \ZI [q^4] ^4 _{t}\right)
  =  -6g_t^2\left(m_t^4 \ZI ^4_t - 12 \ZI [q^4] ^4_{t} + \frac{1}{6}\right).
  \label{eq:delta_Zh_top}
\end{align}
The loop functions that appear in \eqref{eq: HiggsEFT} and
\eqref{eq:delta_Zh_top} can be calculated with the Mathematica file
\texttt{LoopFunctions.m} and read
\begin{align}
  \ZI^1_t &= 2 \ZI[q^2]^2_t = m_t^2 \left(\frac{2}{\eps} + 1 - \log\frac{m_t^2}{\mu^2}\right), \\
  \ZI^2_t &= 24 \ZI[q^4]^4_t = \frac{2}{\eps}-\log\frac{m_t^2}{\mu^2}, \\
  \ZI^4_t &= \frac{1}{6 m_t^4}.
\end{align}

\subsection{MSSM threshold correction to the quartic Higgs coupling}
\label{sec:lambdacalc}

As a first nontrivial application and a check we reproduce the one-loop threshold
correction of the quartic Higgs coupling $\lambda$ when matching the MSSM to the
SM at one-loop \cite{Bagnaschi:2014rsa} in the unbroken phase. As
discussed in \cite{Bagnaschi:2014rsa} there are several contributions
of distinct origins. The scalar contribution
$\Delta \lambda^{1\Loop,\phi}$ arises from interactions of the SM-like
Higgs with heavy Higgs bosons, squarks and sleptons,  and the relevant
interaction Lagrangian is given by
\begin{align}
  \Lag_{\phi} ={}&  - \frac{g_t^2}{2} h^2 (\st{L}^* \st{L} + \st{R}^*\st{R})-\frac{1}{\sqrt{2}} g_t X_t h (\st{L}^* \st{R} + \st{L}\st{R}^*) \nonumber \\ & -\frac{1}{8} c_{2\beta} h^2\sum_{i} \left[\left(g_2^2 - \frac{g_1^2}{5}\right)  \su^* _{Li} \su _{Li}+ \frac{4}{5} g_1^2 \su_{Ri}^* \su_{Ri}- \left(g_2^2 + \frac{g_1^2}{5}\right) \sd_{Li}^* \sd_{Li}- \frac{2}{5} g_1^2 \sd_{Ri}^* \sd_{Ri}\right]
\nonumber \\ &-\frac{1}{8} c_{2\beta} h^2\sum_{i} \left[\left(g_2^2 + g_1^2\frac{3}{5}\right)  \sneu^* _{Li} \sneu _{Li}- \left(g_2^2 - g_1^2\frac{3}{5}\right) \sel_{Li}^* \sel_{Li}- \frac{6}{5} g_1^2 \sel_{Ri}^* \sel_{Ri}\right]
\nonumber \\ & +\frac{1}{16} c_{2\beta}^2 \left(\frac{3}{5} g_1^2 + g_2^2\right) h^2 A^2- \frac{1}{8} \left((1 + s_{2\beta}^2) g_2^2 - \frac{3}{5} g_1^2 c_{2\beta}^2\right) h^2 H^{-} H^{+} \nonumber \\ & - \frac{1}{16} \left(\frac{3}{5} g_1^2 + g_2^2\right) (3 s_{2\beta}^2 - 1) h^2 H^2- \frac{1}{8} \left(\frac{3}{5} g_1^2 + g_2^2\right) s_{2\beta} c_{2\beta} h^3 H \nonumber \\ & + \frac{1}{8} \left(\frac{3}{5} g_1^2 + g_2^2\right) s_{2\beta} c_{2\beta} h^2 (G^{-} H^{+} + H^{-} G^{+})+ \frac{1}{8} \left(\frac{3}{5} g_1^2 + g_2^2\right) s_{2\beta} c_{2\beta} h^2 G^0 A.
\end{align}
Here $g_1$ and $g_2$ are the GUT-normalized electroweak gauge
couplings, $X_t$ is the stop mixing parameter, and $g_t = y_t s_\beta$
with $y_t$ being the MSSM top Yukawa coupling and $s_\beta=\sin (\beta)$. The three generations
of left- and right-handed squarks and sleptons are denoted as
$\su_{Li}$, $\su_{Ri}$, $\sd_{Li}$, $\sd_{Ri}$, $\sel_{Li}$,
$\sel_{Ri}$, $\sneu_{Li}$ ($i=1,2,3$), respectively, where
$\st{L} \equiv \su_{L3}$ and $\st{R} \equiv \su_{R3}$ are the left-
and right-handed stops.  Furthermore we have defined
$h=\sqrt{2}\, \re (\mathcal{H}^0)$, where $\mathcal{H}^0$ is the
neutral component of the SM-like Higgs doublet $\mathcal{H}$ related
to the Higgs doublets $H_u$ and $H_d$ through
\begin{align}
  \mathcal{H} = - c_\beta \epstensor H^{*}_d + s_\beta H_u,
  \label{eq:rot_H}
\end{align}
where $\epstensor$ is the antisymmetric tensor with
$\epstensor_{12}=1$ and $c_\beta = \cos(\beta)$, $s_{2\beta} = \sin(2\beta)$ and
$c_{2\beta} = \cos(2\beta)$.  The fields $G^0$ and $G^{\pm}$ are
Goldstone bosons arising from the same Higgs doublet. The heavy Higgs
bosons $H$, $A$ and $H^\pm$ arise from the heavy doublet
$\mathcal{A}$, which is related to the MSSM doublets through
\begin{align}
  \mathcal{A} = s_\beta \epstensor H^{*}_d + c_\beta H_u.
  \label{eq:rot_A}
\end{align}
Note, that since we work in the unbroken phase, $\beta$ should not be
regarded as a ratio of vacuum expectation values, but as the
fine-tuned mixing angle which rotates the two MSSM Higgs doublets
$H_u$ and $H_d$ into $\mathcal{H}$ and $\mathcal{A}$ as given in
\eqref{eq:rot_H}--\eqref{eq:rot_A} \cite{Bagnaschi:2014rsa}.
The fermionic contribution $\Delta \lambda ^{1\Loop,\chi}$ to the
threshold correction of $\lambda$ originates from interactions of the
Higgs boson with charginos $\tilde{\chi}^{+}_i$ ($i=1,2$) and
neutralinos $\tilde{\chi}^0_i$ ($i=1,\ldots,4$) described by the
interaction Lagrangian
\begin{align}
\Lag_\chi ={}& - \frac{g_2}{\sqrt{2}} h c_\beta (\overline{\tilde{\chi}^{+}_1} P_R \tilde{\chi}^{+} _2  + \overline{\tilde{\chi}^{+}_2}  P_L \tilde{\chi}^{+} _1)- \frac{g_2}{\sqrt{2}} h s_\beta (\overline{\tilde{\chi}^{+}_2} P_R \tilde{\chi}^{+} _1 + \overline{\tilde{\chi}^{+}_1} P_L \tilde{\chi}^{+}_2   )\nonumber 
 \\ &  +i  \frac{g_Y}{2\sqrt{2}} (c_\beta - s_\beta) h \overline{\tilde{\chi}^0_1} \gamma^5  \tilde{\chi}^0_3-\frac{g_Y}{2\sqrt{2}} (c_\beta + s_\beta)h \overline{\tilde{\chi}^0_1} \tilde{\chi}^0_4 \nonumber
 \\ & -i \frac{g_2}{2\sqrt{2}} (c_\beta - s_\beta) h \overline{\tilde{\chi}^0_2} \gamma^5  \tilde{\chi}^0_3+  \frac{g_2}{2\sqrt{2}}  (c_\beta + s_\beta) h \overline{\tilde{\chi}^0_2} \tilde{\chi}^0_4 ,
\end{align}
where $\overline{\tilde{\chi}^0_i} = (\tilde{\chi}^0_i)^T \cc$ and
$g_Y = \sqrt{3/5}\, g_1$.

To calculate the one-loop threshold correction for $\lambda$, the
following contributions with purely scalar and purely fermionic
operators from our generic UOLEA \eqref{eq:UOLEA_final} are relevant,
\begin{align}
\frac{1}{\kappa} \Lag_\EFT^{1\Loop} = \tr \Bigg\lbrace & \frac{1}{2} \ZI ^{1} _{i} (\mathbf{X}_{\Phi \Phi})_{ii}+\frac{1}{2} \ZI [q^2]^{22} _{ij} [P_\mu, (\mathbf{X}_{\Phi \Phi})_{ij}] [P^\mu, (\mathbf{X}_{\Phi \Phi})_{ji}]\nonumber \\ & +\frac{1}{4} \ZI ^{11} _{ij} (\mathbf{X}_{\Phi \Phi})_{ij}(\mathbf{X}_{\Phi \Phi})_{ji}+\frac{1}{6} \ZI^{111} _{ijk}(\mathbf{X}_{\Phi \Phi})_{ij}(\mathbf{X}_{\Phi \Phi})_{jk}(\mathbf{X}_{\Phi \Phi})_{ki} 
\nonumber \\ &+\frac{1}{8} \ZI ^{1111} _{ijkl} (\mathbf{X}_{\Phi \Phi})_{ij}(\mathbf{X}_{\Phi \Phi})_{jk} (\mathbf{X}_{\Phi \Phi})_{kl} (\mathbf{X}_{\Phi \Phi})_{li} + \frac{1}{2} \ZI ^{1} _{i} (\mathbf{X}_{\Phi \phi})_{ij}(\mathbf{X}_{\phi \Phi})_{ji} \nonumber \\ &  -\frac{1}{8} m_{\Xi i}m_{\Xi j} m_{\Xi k} m_{\Xi l} \ZI ^{1111} _{ijkl}(\mathbf{X}_{\Xi \Xi})_{ij}(\mathbf{X}_{\Xi \Xi})_{jk} (\mathbf{X}_{\Xi \Xi})_{kl} (\mathbf{X}_{\Xi \Xi})_{li}
\nonumber \\ &  -\frac{1}{2} m_{\Xi i}m_{\Xi j} \ZI [q^2] ^{1111} _{ijkl}(\mathbf{X}_{\Xi \Xi})_{ij}(\mathbf{X}_{\Xi \Xi})_{jk}\gamma^\mu (\mathbf{X}_{\Xi \Xi})_{kl} \gamma_\mu (\mathbf{X}_{\Xi \Xi})_{li}
\nonumber \\ &  -\frac{1}{4} m_{\Xi i}m_{\Xi k} \ZI [q^2] ^{1111} _{ijkl}(\mathbf{X}_{\Xi \Xi})_{ij}\gamma^\mu(\mathbf{X}_{\Xi \Xi})_{jk} (\mathbf{X}_{\Xi \Xi})_{kl} \gamma_\mu (\mathbf{X}_{\Xi \Xi})_{li}
\nonumber \\ &  -\frac{1}{8} g_{\mu \nu \rho \sigma} \ZI [q^4] ^{1111} _{ijkl}\gamma^\mu (\mathbf{X}_{\Xi \Xi})_{ij}\gamma^\nu(\mathbf{X}_{\Xi \Xi})_{jk} \gamma^\rho (\mathbf{X}_{\Xi \Xi})_{kl} \gamma^\sigma (\mathbf{X}_{\Xi \Xi})_{li}
\nonumber \\ &  +\frac{1}{4} m_{\Xi i} m_{\Xi j}^3 \ZI ^{13} _{ij} [P_\mu,(\mathbf{X}_{\Xi \Xi})_{ij}][P^\mu,(\mathbf{X}_{\Xi \Xi})_{ji}]
\nonumber \\ &  -\frac{1}{2} \ZI[q^4] ^{22} _{ij} \gamma^\nu [P_\mu,(\mathbf{X}_{\Xi \Xi})_{ij}]\gamma_\nu[P^\mu,(\mathbf{X}_{\Xi \Xi})_{ji}]
\nonumber \\ &  - \ZI[q^4] ^{22} _{ij} \gamma^\nu [P_\nu,(\mathbf{X}_{\Xi \Xi})_{ij}]\gamma_\mu[P^\mu,(\mathbf{X}_{\Xi \Xi})_{ji}]
\Bigg\rbrace,
\label{eq:UOLEAToLambda}
\end{align}
where $\kappa=1/(4\pi)^2$.  The operators containing covariant
derivatives can be removed by a field-strength renormalization of the
Higgs field to canonically normalize the kinetic term.  This field
renormalization propagates into every Higgs coupling that has a
non-vanishing tree-level contribution and hence also into the quartic
coupling.  

Next, we compute the $\mathbf{X}_{AB}$ matrices as the
second derivatives of the Lagrangian with respect to the different
kinds of fields.  We start with
\begin{align}
\mathbf{X}_{\Phi \Phi}=\begin{pmatrix} 
X_{\Sigma ^* \Sigma} && X_{\Sigma ^{*} \Sigma ^{*}} &&  X_{\Sigma^* \Theta} \\
X_{\Sigma \Sigma} && X_{\Sigma \Sigma ^{*}} &&  X_{\Sigma \Theta} \\
X_{\Theta \Sigma} && X_{\Theta \Sigma ^*} && X_{\Theta \Theta}\end{pmatrix}
\label{eq: heavy-scalar-heavy-scalar}
\end{align}
and define
\begin{align}
\Sigma &=
\begin{pmatrix}
\su_{Li} & \su_{Ri} & \sd_{Li} & \sd_{Ri} & \sel_{Li} & \sel_{Ri} & \sneu_{Li} & H^{+} 
\end{pmatrix}^T, &
\Theta &= \begin{pmatrix}
A & H
\end{pmatrix}^T ,
\end{align}
where $i=1,2,3$ denotes the generation index.  The non-vanishing
derivatives with respect to two heavy scalar fields read
\begin{align}
X_{\su_{Li}^* \su_{Lj}}&=X_{\su_{Li} \su_{Lj}^*}=\frac{1}{8}c_{2\beta}h^2 \delta_{ij}\left(g_2^2-\frac{1}{5}g_1^2\right)+\delta_{3i}\delta_{3j}\frac{g_t^2}{2}h^2, \\ 
X_{\su_{Ri}^* \su_{Rj}}&=X_{\su_{Ri} \su_{Rj}^*}=\frac{1}{10}c_{2\beta}h^2 \delta_{ij}g_1^2+\delta_{3i}\delta_{3j}\frac{g_t^2}{2}h^2, \\ 
X_{\sd_{Li}^* \sd_{Lj}}&=X_{\sd_{Li} \sd_{Lj}^*}=-\frac{1}{8}c_{2\beta}h^2 \delta_{ij}\left(g_2^2+\frac{1}{5}g_1^2\right), \\ 
X_{\sd_{Ri}^* \sd_{Rj}}&=X_{\sd_{Ri} \sd_{Rj}^*}=\frac{1}{20}c_{2\beta}h^2 \delta_{ij}g_1^2, \\
X_{\sel_{Li}^* \sel_{Lj}}&=X_{\sel_{Li} \sel_{Lj}^*}=\frac{1}{8}c_{2\beta}h^2 \delta_{ij}\left(g_2^2-\frac{3}{5}g_1^2\right), \\  
X_{\sel_{Ri}^* \sel_{Rj}}&=X_{\sel_{Ri} \sel_{Rj}^*}=-\frac{1}{20}c_{2\beta}h^2 \delta_{ij}g_1^2, \\
X_{\sneu_{Li}^* \sneu_{Lj}}&=X_{\sneu_{Li} \sneu_{Lj}^*}=\frac{1}{8}c_{2\beta}h^2 \delta_{ij}\left(g_2^2+\frac{3}{5}g_1^2\right), \\  
X_{H^+ H^-}&=X_{H^- H^+}=\frac{1}{8}h^2 \left[(1+s_{2\beta}^2)g_2^2-\frac{3}{5}g_1^2 c_{2\beta}^2\right] \\
X_{AA}&=-\frac{1}{16}c_{2\beta}^2\left(\frac{3}{5}g_1^2+g_2^2\right)h^2, \\
X_{HH}&=\frac{1}{16}(2s_{2\beta}^2-1)\left(\frac{3}{5}g_1^2+g_2^2\right)h^2, \\
X_{\su_{Li}^* \su_{Rj}}&=X_{\su_{Li} \su_{Rj}^*}=\frac{1}{\sqrt{2}}\delta_{3i}\delta_{3j}g_t X_t h.
\end{align}
Given these derivatives we find that $\mathbf{X}_{\Phi \Phi}$ is
block-diagonal with the blocks being
\begin{align}
X_{\Sigma^* \Sigma}&=\begin{pmatrix}
X_{\su_{Li}^* \su_{Lj}} & X_{\su_{Li}^* \su_{Rj}} & \mathbf{0}_{1\times 6} \\
X_{\su_{Ri}^* \su_{Lj}} & X_{\su_{Ri}^* \su_{Rj}} & \mathbf{0}_{1\times 6} \\
\mathbf{0}_{6\times 1} & \mathbf{0}_{6\times 1} & X_{\Pi^* \Pi} 
\end{pmatrix}, \\
X_{\Pi^* \Pi}&=\diag(X_{\sd_{Li}^* \sd_{Lj}},X_{\sd_{Ri}^* \sd_{Rj}},X_{\sel_{Li}^* \sel_{Lj}},X_{\sel_{Ri}^* \sel_{Rj}},X_{\sneu_{Li}^* \sneu_{Lj}},X_{H^+ H^-}), \\
X_{\Sigma \Sigma^*}&=\begin{pmatrix}
X_{\su_{Li} \su_{Lj}^*} & X_{\su_{Li} \su_{Rj}^*} & \mathbf{0}_{1\times 6} \\
X_{\su_{Ri} \su_{Lj}^*} & X_{\su_{Ri} \su_{Rj}^*} & \mathbf{0}_{1\times 6} \\
\mathbf{0}_{6\times 1} & \mathbf{0}_{6\times 1} & X_{\Pi \Pi^*} 
\end{pmatrix}, \\
X_{\Pi \Pi^*}&=\diag(X_{\sd_{Li} \sd_{Lj}^*},X_{\sd_{Ri} \sd_{Rj}^*},X_{\sel_{Li} \sel_{Lj}^*},X_{\sel_{Ri} \sel_{Rj}^*},X_{\sneu_{Li} \sneu_{Lj}^*},X_{H^- H^+}), \\
X_{\Theta \Theta}&=\diag(X_{AA},X_{HH}),
\end{align}
where $\mathbf{0}_{m\times n}$ denotes the $m \times n$ matrix of only
zeros. We next calculate $\mathbf{X}_{\phi \Phi}$ and
$\mathbf{X}_{\Phi\phi}$, which contain derivatives with respect to one
heavy and one light scalar field.  We define the light scalar field
multiplets as
\begin{align}
  \sigma &= (G^+), &
  \theta &=
           \begin{pmatrix}
             h & G^0
           \end{pmatrix}^T.
\end{align}
As discussed in \secref{sec: intro} the derivatives w.r.t.\ the fields
are evaluated at the background field configurations, and the heavy
background fields are expressed in terms of the light ones using a
local operator expansion.\footnote{An explicit example is given in
  \secref{sec: gluinoOut} in the treatment of dimension 5
  operators.} This corresponds to an expansion in $\Box/M^2$ for a
heavy scalar field of mass $M$ and hence it leads to contributions
suppressed by at least $1/M^2$.  Since we are not interested in these
suppressed contributions here, we only consider derivatives of the
Lagrangian which exclusively contain light background fields and set
all other derivatives to zero. The non-vanishing derivatives are given
by
\begin{align}
X_{Hh} &= X_{hH}=\frac{3}{8}\left(\frac{3}{5}g_1^2+g_2^2\right)s_{2\beta}c_{2\beta}h^2 ,\\
X_{AG^0} &= X_{G^0A}=-\frac{1}{8}\left(\frac{3}{5}g_1^2+g_2^2\right)s_{2\beta}c_{2\beta}h^2, \\
X_{H^{+}G^{-}} &= X_{H^{-}G^{+}}=-\frac{1}{8}\left(\frac{3}{5}g_1^2+g_2^2\right)s_{2\beta}c_{2\beta}h^2.
\end{align}
We then find that $\mathbf{X}_{\Phi \phi}$ is block-diagonal with the blocks being
\begin{align}
X_{\Sigma^* \sigma}&=\begin{pmatrix} \mathbf{0}_{7 \times 1} \\ X_{H^{-} G^{+}}
\end{pmatrix}, \\
X_{\Sigma \sigma^*}&=\begin{pmatrix} \mathbf{0}_{7 \times 1} \\ X_{H^{+} G^{-}}
\end{pmatrix}, \\
X_{\Theta \theta}&=\begin{pmatrix}  0 & X_{A G^0}\\
X_{Hh} & 0
\end{pmatrix}.
\end{align}
Similarly, $\mathbf{X}_{\phi \Phi}$ is block-diagonal with diagonal entries
\begin{align}
X_{\sigma^* \Sigma}&=\begin{pmatrix} \mathbf{0}_{1 \times 7} & X_{G^{-} H^{+}}
\end{pmatrix}, \\
X_{\sigma \Sigma^*}&=\begin{pmatrix} \mathbf{0}_{1 \times 7} & X_{G^{+} H^{-}}
\end{pmatrix}, \\
X_{\theta \Theta}&=\begin{pmatrix}   0 & X_{h H}\\
 X_{G^0 A} & 0
\end{pmatrix}.
\end{align}
Finally, we need the derivatives with respect to two heavy fermions to
construct the matrix $\mathbf{X}_{\Xi \Xi}$.  We define
\begin{align}
\Omega &= \begin{pmatrix}
\tilde{\chi}^+ _1 & \tilde{\chi}^+ _2
\end{pmatrix}^T, & \Lambda &= \begin{pmatrix}
\tilde{\chi}^0 _1 & \tilde{\chi}^0 _2 & \tilde{\chi}^0 _3 & \tilde{\chi}^0 _4
\end{pmatrix}^T
\end{align}
and the matrix $\mathbf{X}_{\Xi \Xi}$ is again block-diagonal with the
non-vanishing entries
\begin{align}
X_{\bar{\Omega} \Omega}&=\cc ^{-1} X^T_{\Omega \bar{\Omega}} \cc^{-1}=-\frac{g_2}{\sqrt{2}}h\begin{pmatrix}
0 & c_\beta P_R+ s_\beta P_L \\
c_\beta P_L+s_\beta P_R & 0
\end{pmatrix}, \\
\cc^{-1} X_{\Lambda \Lambda}&=\frac{h}{2\sqrt{2}}\begin{pmatrix}
0 & 0 & i g_Y(c_\beta-s_\beta)\gamma^5 & -g_Y(c_\beta+s_\beta) \\
0 & 0 & -ig_2(c_\beta-s_\beta)\gamma^5 & g_2(c_\beta+s_\beta) \\
i g_Y(c_\beta-s_\beta)\gamma^5 & -ig_2(c_\beta-s_\beta)\gamma^5 & 0 & 0 \\
-g_Y(c_\beta+s_\beta) & g_2(c_\beta+s_\beta) & 0 & 0
\end{pmatrix}, 
\end{align}
where the relations of \appref{sec: spinor algebra} were used to
simplify the expressions.  Note, that in the calculation of
$X_{\Lambda \Lambda}$ for a given Majorana fermion $\lambda$ the two
fields $\bar{\lambda}$ and $\lambda$ are not independent, but are
related via $\bar{\lambda}=\lambda^T \cc$. Inserting all of the
derivatives into \eqref{eq:UOLEAToLambda}, summing over all indices
and canonically normalizing the kinetic term for the SM-like Higgs
boson as
\begin{align}
  h ={}& \left(1 - \frac{1}{2} \delta Z_h\right) \hat{h}, \\
  \delta Z_h ={}& -6 g_t^2 X_t^2 \ZI[q^2]_{\sq\su}^{22}+\frac{s_{2 \beta}}{2} \mu \left(g^2_Y M_1 \mu^2 \ZI ^{13}_{1\mu}+g^2_Y M_1^3  \ZI ^{31}_{1\mu}-3g^2_2 M_2 \mu^2 \ZI ^{13}_{2\mu}-3g^2_2 M_2^3  \ZI ^{31}_{2\mu}\right) 
  \nonumber \\& +2(2+d)\left(-g_Y^2 \ZI[q^4]^{22}_{1\mu}+3g_2^2 \ZI[q^4]^{22}_{2\mu}\right) ,
\end{align}
one finds the following effective Lagrangian
\begin{align}
  \Lag_{\EFT}^{1\Loop} = \frac{1}{2}(\partial \hat{h})^2
  % + \frac{m^2}{2} \hat{h}^2
  - \frac{\lambda}{8} \hat{h}^4 + \cdots
\end{align}
with
%
%As an application we consider the MSSM with all SUSY particles heavy,
%and all with masses around the same mass scale \MS.
%
%\begin{align}
%  \begin{split}
%    \LagSM ={}& - G^d_{ij} (H^\dagger q_j)d_i + G^u_{ij} (H^T\epstensor q_j)u_i - G^e_{ij} (H^T\epstensor l_j)e_i + \hc  \\
%    & + m^2 |H|^2 - \frac{\lambda}{2} |H|^4 ,
%  \end{split}
%  \label{eq:LSM}
%\end{align}
%
%with $l_{i} = (\nu_{iL}, e_{iL})^T$, $e_{i} = e_{iR}$,
%$\tilde{l}_{i} = (\tilde{\nu}_{iL}, \tilde{e} _{iL})^T$ and
%$\epstensor_{12} = 1$.  All parameters in \eqref{eq:LSM} are
%renormalized in the \MSbar\ scheme.
%
%The CKM and PMNS matrices are set diagonal and we define
%
%\begin{align}
%  (G^e_{ij}) &= \diag(g_e, g_\mu, g_\tau), \\
%  (G^u_{ij}) &= \diag(g_u, g_c, g_t), \\
%  (G^d_{ij}) &= \diag(g_d, g_s, g_b).
%\end{align}
%
%We fix the quartic Higgs coupling $\lambda$ of \eqref{eq:LSM} at the matching
%scale \MS in terms of SM \MSbar\ and MSSM \DRbarPrime\ parameters as
%
\begin{align}
  \lambda &= \frac{1}{4} \left( \frac{3}{5} g_1^2 + g_2^2 \right) c_{2\beta}^2
           + \kappa \Delta\lambda^{1\Loop} , \\
   \Delta\lambda^{1\Loop} &= \Delta\lambda^{1\Loop,\text{reg}}
           + \Delta\lambda^{1\Loop,\phi} + \Delta\lambda^{1\Loop,\chi},
\end{align}
and
\begin{align}
  \Delta \lambda^{1\Loop,\phi} ={}&
  g_t^4 \left[
     -3 X_t^4 \ZI_{\sq\sq\su\su}^{1111}
     -6 X_t^2 \left(\ZI_{\sq\sq\su}^{111} + \ZI_{\sq\su\su}^{111}\right)
     -3 \left(\ZI_{\sq\sq}^{11} + \ZI_{\su\su}^{11}\right)
   \right] \nonumber \\
   &+\frac{3}{10} g_t^2 c_{2\beta} \Big\{ X_t^2 \left[
   2 c_{2\beta} \left(3 g_1^2+5g_2^2\right) \ZI[q^2]_{\sq\su}^{22}
   +\left(g_1^2-5 g_2^2\right) \ZI_{\sq\sq\su}^{111}
   -4 g_1^2 \ZI_{\sq\su\su}^{111}\right] \nonumber \\
   &~~~~~~~~~~~~~~~~ + \left(g_1^2-5 g_2^2\right) \ZI_{\sq\sq}^{11}
   -4 g_1^2 \ZI_{\su\su}^{11}\Big\} \nonumber \\
   & -\frac{c_{2\beta}^2}{200} \sum_{i=1}^3 \Big[
   3 \left(g_1^4+25 g_2^4\right) \ZI_{\sq_i\sq_i}^{11}
   +24 g_1^4 \ZI_{\su_i\su_i}^{11}
   +6 g_1^4 \ZI_{\sd_i\sd_i}^{11} \nonumber \\
   &~~~~~~~~~~~~~~~ +\left(9 g_1^4+25 g_2^4\right) \ZI_{\slep_i\slep_i}^{11}
   +18 g_1^4 \ZI_{\sel_i\sel_i}^{11}
   \Big] \nonumber \\
   &+ \frac{1}{200} \Big\{6 c_{2\beta}^2 \left(c_{2\beta}^2-1\right) \left(3
   g_1^2+5 g_2^2\right)^2 \ZI_{A0}^{11} - \Big[9 \left(3
   c_{2\beta}^4-3 c_{2\beta}^2+1\right) g_1^4 \nonumber \\
   &~~~~~~~~~~ +30 \left(3 c_{2\beta}^4-4
   c_{2\beta}^2+1\right) g_1^2 g_2^2+25 \left(3 c_{2\beta}^4-5
   c_{2\beta}^2+3\right) g_2^4\Big]
   \ZI_{AA}^{11}\Big\},\\
  \Delta\lambda^{1\Loop,\chi} ={}& -\frac{1}{4}
  \Big\{-d \big(2 g_Y^4 M_1^2 \ZI[q^2] ^{22} _{1 \mu} + 
      2 g_2^4 M_2^2 \ZI[q^2] ^{22} _{2 \mu} + 
      g_Y^4 \mu^2 \ZI[q^2] ^{22} _{1 \mu} \nonumber \\ &\qquad ~~~~~~~~ - 
      g_Y^4 \mu^2 c_{4 \beta} \ZI[q^2] ^{22} _{1 \mu} +
      g_2^4 \mu^2 \ZI[q^2] ^{22} _{2 \mu} 
      - g_2^4 \mu^2 c_{4 \beta} \ZI[q^2] ^{22} _{2 \mu} \nonumber \\ &\qquad ~~~~~~~~ + 
      4 g_Y^2 g_2^2 M_1 M_2 \ZI[q^2] ^{112} _{1 2 \mu} + 
      2 g_Y^2 g_2^2 \mu^2 \ZI[q^2] ^{112} _{1 2 \mu} - 
      2 g_Y^2 g_2^2 \mu^2 c_{4 \beta} \ZI[q^2] ^{112} _{1 2 \mu}\big) \nonumber \\ &\qquad~~   - 
   d (2 + d) \big(2 g_Y^4 \ZI[q^4] ^{22} _{1 \mu} + 
      2 g_2^4 \ZI[q^4] ^{22} _{2 \mu} + 
      4 g_Y^2 g_2^2 \ZI[q^4] ^{112} _{1 2 \mu}\big) \nonumber \\ &\qquad~~ 
       - g_2^4 \big[2d (2 + d) (3 + c_{4 \beta}) \ZI[q^4] ^{22} _{2 \mu} 
   + 16 c_{\beta} s_{\beta} (d M_2 \ZI[q^2] ^{22} _{2 \mu} (\mu + M_2 c_{\beta} s_{\beta}) 
   \nonumber \\ &\qquad ~~~~~~~~~ +
          \mu \{M_2^2 \mu c_{\beta} \ZI ^{22} _{2 \mu} s_{\beta} + 
            d \ZI[q^2] ^{22} _{2 \mu} (M_2 + \mu c_{\beta} s_{\beta})\})\big]
            \nonumber \\ &\qquad~~  - 
   4 d \mu \big(2 g_Y^4 M_1 \ZI[q^2] ^{22} _{1 \mu} + 
      2 g_2^4 M_2 \ZI[q^2] ^{22} _{2 \mu} + 
      2 g_Y^2 g_2^2 M_1 \ZI[q^2] ^{112} _{1 2 \mu}  \nonumber \\ &\qquad ~~~~~~~~~~~ + 
      2 g_Y^2 g_2^2 M_2 \ZI[q^2] ^{112} _{1 2 \mu}\big) s_{2 \beta} 
      \nonumber \\ &\qquad~~ - 
   2 \mu^2 \big(g_Y^4 M_1^2 \ZI ^{22} _{1 \mu} + 
      g_2^2 M_2 (g_2^2 M_2 \ZI ^{22} _{2 \mu} + 
         g_Y^2 M_1 2\ZI ^{112} _{12 \mu} + 
            )\big) s_{2 \beta}^2 \nonumber \\ &\qquad~~  - 
   2 g_2^2 (g_Y^2 + g_2^2) c_{
     2 \beta}^2 \big(-4 (2 + d) \ZI[q^4] ^{22} _{2 \mu} + 
      M_2 \mu (\mu^2 \ZI ^{13} _{2 \mu} + 
         M_2^2 \ZI ^{31} _{2 \mu}) s_{2 \beta}\big)\nonumber \\ &\qquad~~  - (g_Y^2 + 
      g_2^2) c_{
     2 \beta}^2 \big(-4 (2 + d) g_Y^2 \ZI[q^4] ^{22} _{1 \mu} - 
      4 (2 + d) g_2^2 \ZI[q^4] ^{22} _{2 \mu} \nonumber \\ &\qquad ~~~~~~ + 
      \mu \{g_Y^2 M_1 \mu^2 \ZI ^{13} _{1 \mu} + 
         g_Y^2 M_1^3 \ZI ^{31} _{1 \mu} + 
         g_2^2 M_2 (\mu^2 \ZI ^{31} _{2 \mu} + 
            M_2^2 \ZI ^{31} _{2 \mu})\} s_{2 \beta}\big)\Big\}.
\end{align}
The subscripts $1$ and $2$ of the loop functions are
shorthand for $M_1$ and $M_2$, respectively. The terms involving
$d = 4 - \eps$ originate from contractions of gamma matrices and
metric tensors, see \appref{sec:DREG_DRED}.  Note, that
$\lambda$ is expressed entirely in terms of the MSSM gauge couplings,
in contrast to \cite{Bagnaschi:2014rsa}.

It is sensible to regularize the MSSM using dimensional reduction
(DRED) \cite{Siegel:1979wq}, whereas the SM is more naturally
regularized in dimensional regularization (DREG)
\cite{Bollini:1972ui,Ashmore:1972uj,Cicuta:1972jf,tHooft:1972tcz,tHooft:1973mfk}.
Such a regularization scheme change leads to further contributions to
the threshold correction denoted by
$\Delta\lambda^{1\Loop,\text{reg}}$, which can be obtained using the
DRED--DREG regularization scheme translating operators presented in
\cite{Summ:2018oko}. This contribution originates from the operator
\begin{align}
\frac{1}{\kappa} \eps \Lag_{\EFT,\eps}^{1\Loop} =
 \frac{1}{2} \tr \{\epsdim{X}^{\mu \nu}_{\eps \eps} \epsdim{X}_{\eps \eps \mu \nu} \},
\label{eq: epsilonconttolambda}
\end{align}
where on the r.h.s.\ $\eps$ denotes all epsilon scalars that couple to
the Higgs and
\begin{align}
  \epsdim{X}^{\mu \nu}_{\eps \eps} = \epsdim{g}^\mu_\sigma \epsdim{g}^\nu_\rho \fourdim{X}^{\sigma\rho}_{\eps \eps}
\end{align}
is the projection of the $4$-dimensional
$\fourdim{X}^{\sigma\rho}_{\eps \eps}$ onto the
$\eps$-dimensional $Q\eps S$ space
\cite{Stockinger:2005gx,Summ:2018oko} with
$\epsdim{g}^{\mu\nu}\epsdim{g}_{\mu\nu} = \eps$, see
\appref{sec:DREG_DRED}.
In the MSSM we have the following couplings to epsilon scalars to the
SM-like doublet $\mathcal{H}$,
\begin{align}
\Lag_{\eps \mathcal{H}}=\mathcal{H}^*_i \epsdim{g}_{\mu \nu}\left(g_2 ^2 T^a_{ij} T^b_{jl} a^{a\mu} a^{b\nu}+\sqrt{\frac{3}{5}} g_1 g_2 T^a_{il} a^{a\mu} b^\nu +\frac{3}{20}g_1^2 b^\mu b^\nu \delta_{il}\right)\mathcal{H}_l,
\end{align}
where the indices $i,j,l$ are $SU(2)_L$ indices of the fundamental
representation with the generators $T^a_{ij}$.  The fields
$a^{a \mu}$ and $b^\mu$ denote the epsilon scalars
corresponding to $SU(2)_L$ and $U(1)_Y$, respectively.  One obtains
the derivative
\begin{align}
\epsdim{X}^{\mu \nu}_{\eps \eps}=-\epsdim{g}^{\mu \nu} \begin{pmatrix}
\mathcal{H}^{*}_i g_2^2 \{T^a,T^b\}_{il} \mathcal{H}_l & \sqrt{\frac{3}{5}} g_1 g_2 \mathcal{H}^{*}_i T^a _{il} \mathcal{H}_l \\
\sqrt{\frac{3}{5}} g_1 g_2 \mathcal{H}^{*}_i T^a _{il} \mathcal{H}_l & \frac{3}{10} g_1^2 \mathcal{H}^{*}_i \mathcal{H}_i  
\end{pmatrix}.
\end{align}
Inserting this into \eqref{eq: epsilonconttolambda} we obtain
\begin{align}
\Delta\lambda^{1\Loop,\text{reg}} &=
      - \frac{9}{100}g_1^4 - \frac{3}{10} g_1^2 g_2^2 
      -\frac{3}{4} g_2^4.
\end{align}
We do not find the term proportional to $c_{2\beta}^2$ given in
\cite{Bagnaschi:2014rsa} since this term only arises once the
tree-level expression for $\lambda$ is expressed in terms of SM gauge
couplings, as opposed to MSSM parameters as in our case. Up to terms
arising from this conversion the one-loop threshold corrections agree
with the results of \cite{Bagnaschi:2014rsa}.

\subsection{Integrating out stops and the gluino from the MSSM}
\label{sec:matching_MSSM_to_SMEFT}

As a second nontrivial application we reproduce known threshold corrections from
the MSSM to the Standard Model Effective Field Theory (SMEFT) from
heavy stops and the gluino in the gaugeless limit ($g_1 = g_2 = 0$) in
the unbroken phase and for vanishing Yukawa couplings, except for the
one of the top quark.  In particular we reproduce the Wilson
coefficient of the higher-dimensional $\hat{h}^6$ operator calculated
in \cite{Drozd:2015rsp,Bagnaschi:2017xid}.  Furthermore, this example
application again represents a scenario, where a heavy Majorana
fermion is integrated out and the formalism introduced in
\secref{sec:calculation} must be carefully applied.

We consider the following part of the MSSM Lagrangian
\begin{align}
  \begin{split}
  \Lag_\MSSM \supset{}&
  |\partial\st{L}|^2 - \mstL |\st{L}|^2
  + |\partial\st{R}|^2 - \mstR |\st{R}|^2
  + \frac{1}{2}(\gluino{a})^T \cc (i\slashed{\partial} - m_{\gluino{}}) \gluino{a}\\
  &
  - \frac{y_t s_\beta}{\sqrt{2}} h \bar{t} t
  - \frac{y_t^2 s_\beta^2}{2} h^2 \left(|\st{L}|^2 + |\st{R}|^2\right)
  - \frac{y_t s_\beta X_t}{\sqrt{2}} h \left(\st{L}^* \st{R} + \hc\right) \\
  &
  - \sqrt{2} g_3 \left[
    \bar{t} P_R \gluino{a} T^a \st{L} - \bar{t} P_L \gluino{a} T^a \st{R}
    + \st{L}^* (\gluino{a})^T T^a \cc P_L t - \st{R}^* (\gluino{a})^T T^a \cc P_R t
  \right] ,
  \end{split}
  \label{eq:LMSSM_stop}
\end{align}
where we use the same notation as in \secref{sec:lambdacalc} and $g_3$
is the strong gauge coupling.  The top
quark is denoted as $t$ and is defined as a Dirac fermion built from
the upper component of the left-handed quark-doublet $q_L$ and the
right-handed top $t_R$.  The gluino is denoted as $\gluino{a}$ and we
have used the relation
$\overline{\gluino{a}} = (\ccfield{(\gluino{a})})^T \cc =
(\gluino{a})^T \cc$ to express \eqref{eq:LMSSM_stop} in terms of the
gluino Majorana spinor $\gluino{a}$.

Upon integrating out the heavy stops and the gluino the Lagrangian of
the effective theory becomes
\begin{align}
  \Lag_\SMEFT \supset
  - \frac{y_t s_\beta}{\sqrt{2}} h \bar{t} t
  + \Lag_\SMEFT^\text{1\Loop}.
\end{align}
In our limit the one-loop term $\Lag_\SMEFT^\text{1\Loop}$ receives
contributions from the following generic operators from
\eqref{eq:UOLEA_final}
\begin{align}
  \frac{1}{\kappa} \Lag_\EFT^{1\Loop} \supset{}& \frac{1}{2} \ZI ^1 _i  (\mathbf{X}_{\Phi \Phi})_{ii}+\frac{1}{4} \ZI ^{11} _{ik} (\mathbf{X}_{\Phi \Phi})_{ik} (\mathbf{X}_{\Phi \Phi})_{ki}+\frac{1}{6} \ZI ^{111} _{lik} (\mathbf{X}_{\Phi \Phi})_{ik} (\mathbf{X}_{\Phi \Phi})_{kl} (\mathbf{X}_{\Phi \Phi})_{li}\nonumber \\
  & +\frac{1}{8} \ZI ^{1111} _{likn} (\mathbf{X}_{\Phi \Phi})_{ik} (\mathbf{X}_{\Phi \Phi})_{kl} (\mathbf{X}_{\Phi \Phi})_{ln} (\mathbf{X}_{\Phi \Phi})_{ni}\nonumber \\ 
  & +\frac{1}{10} \ZI ^{11111} _{iklnp} (\mathbf{X}_{\Phi \Phi})_{ik} (\mathbf{X}_{\Phi \Phi})_{kl} (\mathbf{X}_{\Phi \Phi})_{ln} (\mathbf{X}_{\Phi \Phi})_{np} (\mathbf{X}_{\Phi \Phi})_{pi}\nonumber \\
  & +\frac{1}{12} \ZI ^{111111} _{iklnpr} (\mathbf{X}_{\Phi \Phi})_{ik} (\mathbf{X}_{\Phi \Phi})_{kl} (\mathbf{X}_{\Phi \Phi})_{ln} (\mathbf{X}_{\Phi \Phi})_{np} (\mathbf{X}_{\Phi \Phi})_{pr} (\mathbf{X}_{\Phi \Phi})_{ri}\nonumber \\
  & +\frac{1}{2} \ZI [q^2]^{22} _{ki} [P_\mu, (\mathbf{X}_{\Phi \Phi})_{ik}] [P^\mu, (\mathbf{X}_{\Phi \Phi})_{ki}]\nonumber \\
  & -\ZI [q^2] ^{21}_{il} (\mathbf{X}_{\Phi \Xi})_{il} \gamma^\mu [P_\mu, (\mathbf{X}_{\Xi \Phi})_{li}]\nonumber \\
  & - \frac{1}{2} m_{\Xi_{k}} \ZI ^{111} _{ikl} (\mathbf{X}_{\Phi \Xi})_{ik} (\mathbf{X}_{\Xi \Phi})_{kl} (\mathbf{X}_{\Phi \Phi})_{li} .
\label{eq:L_SMEFT_operators}
\end{align}
We furthermore
set $P_\mu \equiv i \partial_\mu$ to omit contributions from gauge
bosons.  In our scenario we identify $\Sigma = (\st{L}, \st{R})$ as
the vector of (complex) heavy stops and $\Lambda = \gluino{a}$ as the
heavy gluino.  From \eqref{eq:LMSSM_stop} we then obtain the following
non-vanishing derivatives
\begin{align}
  (X_{\st{L}^* \st{L}})_{ij} &= (X_{\st{L} \st{L}^*})_{ij} =
  (X_{\st{R}^* \st{R}})_{ij} = (X_{\st{R} \st{R}^*})_{ij} = \frac{1}{2}(y_t s_\beta h)^2 \delta_{ij},\\
  (X_{\st{L}^* \st{R}})_{ij} &= (X_{\st{L} \st{R}^*})_{ij} =
  (X_{\st{R}^* \st{L}})_{ij} = (X_{\st{R} \st{L}^*})_{ij} = \frac{1}{\sqrt{2}} y_t s_\beta h X_t \delta_{ij},\\
  (X_{\st{L} \gluino{a}})_{i\alpha}^a &= (X_{\gluino{a} \st{L}})_{i\alpha}^a = -\sqrt{2} g_3 (\bar{t}_j P_R)_{\alpha} T^a_{ji}, \label{eq:sign_1}\\
  (X_{\st{R} \gluino{a}})_{i\alpha}^a &= (X_{\gluino{a} \st{R}})_{i\alpha}^a = \sqrt{2} g_3 (\bar{t}_j P_L)_{\alpha} T^a_{ji}, \label{eq:sign_2}\\
  (X_{\gluino{a} \st{L}^*})_{i\alpha}^a &= (X_{\st{L}^* \gluino{a}})_{i\alpha}^a = \sqrt{2} g_3 T^a_{ij} (\cc P_L t_j)_{\alpha},\\
  (X_{\gluino{a} \st{R}^*})_{i\alpha}^a &= (X_{\st{R}^* \gluino{a}})_{i\alpha}^a =  -\sqrt{2} g_3 T^a_{ij} (\cc P_R t_j)_{\alpha},
\end{align}
where $i,j=1,2,3$ and $a=1,\ldots,8$ are color indices and
$\alpha=1,\ldots,4$ is a 4-component spinor index.  Note the flipped
sign in eqs.\ \eqref{eq:sign_1}--\eqref{eq:sign_2} due to one
anti-commutation of the spinor $\bar{t}$ with the derivative w.r.t.\
the spinor $\gluino{a}$.  The bold derivative matrices thus become
\begin{align}
  \mathbf{X}_{\Phi \Phi} &=
  \begin{pmatrix} 
    X_{\Sigma ^* \Sigma} & X_{\Sigma ^* \Sigma ^*} \\
    X_{\Sigma \Sigma} & X_{\Sigma \Sigma ^{*}}
  \end{pmatrix} =
  \begin{pmatrix}
    (X_{\st{L}^* \st{L}})_{ij} & (X_{\st{L}^* \st{R}})_{ij} & 0 & 0 \\
    (X_{\st{R}^* \st{L}})_{ij} & (X_{\st{R}^* \st{R}})_{ij} & 0 & 0 \\
    0 & 0 & (X_{\st{L} \st{L}^*})_{ij} & (X_{\st{L} \st{R}^*})_{ij} \\
    0 & 0 & (X_{\st{R} \st{L}^*})_{ij} & (X_{\st{R} \st{R}^*})_{ij}
  \end{pmatrix} \\
  &= \delta_{ij} \; \mathbf{1}_{2\times 2} \otimes
  \begin{pmatrix}
    \frac{1}{2}(y_t s_\beta h)^2 & \frac{1}{\sqrt{2}} y_t s_\beta h X_t \\
    \frac{1}{\sqrt{2}} y_t s_\beta h X_t & \frac{1}{2}(y_t s_\beta h)^2
  \end{pmatrix}, \\
  \mathbf{X}_{\Phi \Xi} &=
  \begin{pmatrix}
    X_{\Sigma ^* \Lambda} \\
    X_{\Sigma \Lambda}
  \end{pmatrix}
  =
  \begin{pmatrix}
    (X_{\st{L}^* \gluino{a}})_{i\alpha}^a \\
    (X_{\st{R}^* \gluino{a}})_{i\alpha}^a \\
    (X_{\st{L} \gluino{a}})_{i\alpha}^a \\
    (X_{\st{R} \gluino{a}})_{i\alpha}^a
  \end{pmatrix}
  = \sqrt{2} g_3
  \begin{pmatrix}
    T^a_{ij} (\cc P_L t_j)_{\alpha} \\
     -T^a_{ij} (\cc P_R t_j)_{\alpha} \\
     -(\bar{t}_j P_R)_{\alpha} T^a_{ji} \\
    (\bar{t}_j P_L)_{\alpha} T^a_{ji}
  \end{pmatrix},
  \\
  \mathbf{X}_{\Xi \Phi} &=
  \begin{pmatrix}
    \cc^{-1} X_{\Lambda \Sigma}, && \cc^{-1} X_{\Lambda \Sigma ^*}
  \end{pmatrix} \\
  &= (\cc^{-1})_{\alpha\beta}
  \begin{pmatrix}
    (X_{\gluino{a} \st{L}})_{i\beta}^a, &&
    (X_{\gluino{a} \st{R}})_{i\beta}^a, &&
    (X_{\gluino{a} \st{L}^*})_{i\beta}^a, &&
    (X_{\gluino{a} \st{R}^*})_{i\beta}^a
  \end{pmatrix} \\
  &= \sqrt{2} g_3 (\cc^{-1})_{\alpha\beta}
  \begin{pmatrix}
    -(\bar{t}_j P_R)_{\beta} T^a_{ji}, &&
   (\bar{t}_j P_L)_{\beta} T^a_{ji}, &&
   T^a_{ij} (\cc P_L t_j)_{\beta}, &&
    -T^a_{ij} (\cc P_R t_j)_{\beta}
  \end{pmatrix} \\
  &= \sqrt{2} g_3
  \begin{pmatrix}
    -(\bar{t}_j P_R (\cc^{-1})^T)_{\alpha} T^a_{ji}, &&
   (\bar{t}_j P_L (\cc^{-1})^T)_{\alpha} T^a_{ji}, &&
   T^a_{ij} (P_L t_j)_{\alpha}, &
    -T^a_{ij} (P_R t_j)_{\alpha}
  \end{pmatrix} .
\end{align}
By inserting the $\mathbf{X}_{AB}$ operators into
\eqref{eq:L_SMEFT_operators} and summing over all fields and colors we
obtain
\begin{align}
  \Lag_\EFT^\text{1\Loop} &=
     c_t h\bar{t}t + c_L\bar{t}i\slashed{\partial}P_Lt + c_R \bar{t}i\slashed{\partial}P_Rt
     + c_2' (\partial h)^2 + c_2 h^2 + c_4 h^4 + c_6 h^6 + \cdots,
\end{align}
where
\begin{align}
  c_t &= -\frac{4 \sqrt{2}}{3}\kappa g_3^2 y_t s_\beta m_{\gluino{}} X_t \ZI ^{111} _{\gluino{}\sq\su},\\
\begin{split}
  c_L &= \frac{16}{3}\kappa g_3^2 \ZI [q^2] ^{21} _{\su \gluino{}},
\end{split}\\
  c_R &= c_L|_{\sq \to \su},\\
  c_2' &= -3 \kappa (y_t s_\beta)^2 X_t^2 \ZI [q^2] ^{22} _{\sq \su},\\
  c_2 &= \frac{3}{2}\kappa (y_t s_\beta)^2 \left[\ZI ^{1} _{\sq} + \ZI ^{1} _{\su} + X_t^2 \ZI ^{11} _{\sq\su}\right], \\
  c_4 &= \frac{3}{8}\kappa (y_t s_\beta)^4 \left[
         \ZI ^{11} _{\sq\sq} + \ZI ^{11} _{\su\su} + 2 X_t^2 (\ZI ^{111} _{\sq\sq\su} + \ZI ^{111} _{\sq\su\su}) + X_t^4 \ZI ^{1111} _{\sq\sq\su\su}\right],\\
\begin{split}
  c_6 &= \frac{1}{8}\kappa (y_t s_\beta)^6 \big[
         \ZI ^{111} _{\sq\sq\sq} + \ZI ^{111} _{\su\su\su} + 3 X_t^2 ( \ZI ^{1111} _{\sq\sq\sq\su} + \ZI ^{1111} _{\sq\sq\su\su} + \ZI ^{1111} _{\sq\su\su\su} ) \\
    & ~~~~~~~~~~~~~~~~~~
    + 3 X_t^4 ( \ZI ^{11111} _{\sq\sq\sq\su\su} + \ZI ^{11111} _{\sq\sq\su\su\su} )
    + X_t^6 \ZI ^{111111} _{\sq\sq\sq\su\su\su} \big] .
\end{split}
\end{align}
To canonically normalize the kinetic terms
of $\Lag_\SMEFT$ we re-define the Higgs and the top quark field as
\begin{align}
  h  &= \left(1 - \frac{1}{2} \delta Z_h\right) \hat{h} , \\
  t_L &= \left(1 - \frac{1}{2} \delta Z_L\right) \hat{t}_L , \\
  t_R &= \left(1 - \frac{1}{2} \delta Z_R\right) \hat{t}_R ,
\end{align}
where the field renormalizations $\delta Z_{h/L/R}$ are given by
\begin{align}
  \delta Z_h &= 2c_2', \\
  \delta Z_L &= c_L, \\
  \delta Z_R &= c_R.
\end{align}
If we parameterize the \SMEFT Lagrangian as
\begin{align}
  \Lag_\SMEFT \supset
  - \frac{g_t}{\sqrt{2}} \hat{h} \bar{\hat{t}} \hat{t}
  + \frac{m^2}{2} \hat{h}^2 - \frac{\lambda}{8} \hat{h}^4
  - \frac{\tilde{c}_6}{8} \hat{h}^6,
\end{align}
then the SMEFT parameters $g_t$, $\lambda$ and $m^2$ are given by
\begin{align}
  g_t &= y_t s_\beta \left[1 - \frac{1}{2}(c_L + c_R) - c_2' - \frac{\sqrt{2}c_t}{y_t s_\beta} \right], \\
  m^2 &= 2 c_2,\\
  \lambda &= -8 c_4 ,\\
  \tilde{c}_6 &= -8 c_6,
\end{align}
which agrees with the results calculated in
\cite{Bagnaschi:2014rsa,Bagnaschi:2017xid,Huo:2015nka,Drozd:2015rsp}.\footnote{It
  was noted in \cite{Bagnaschi:2017xid} that the logarithmic term in
  the last line of eq.~(D.4) in \cite{Drozd:2015rsp} should come with a
  minus sign.}

\subsection{Integrating out the gluino from the MSSM with light stops}
\label{sec: gluinoOut}

In this section we calculate some of the terms that arise when
integrating out the gluino from the MSSM.  This \EFT\ scenario is
relevant when there is a large hierarchy between the gluino mass and
the stop masses in the MSSM.  This example is also a direct
application of most of the operators calculated in \secref{sec:calc},
in particular operators where Majorana and Dirac fermions appear in
loops at the same time.

We consider the following part of the MSSM Lagrangian
\begin{align}
 \Lag_\MSSM \supset{}&
  |\partial\st{L}|^2 - \mstL |\st{L}|^2
  + |\partial\st{R}|^2 - \mstR |\st{R}|^2
  + \frac{1}{2}(\gluino{a})^T \cc (i\slashed{\partial} - m_{\gluino{}}) \gluino{a} \nonumber \\
  & -\sqrt{2} g_3 \left(
    \bar{t} P_R \gluino{a} T^a \st{L} - \bar{t} P_L \gluino{a} T^a \st{R}
    + \st{L}^* (\gluino{a})^T T^a \cc P_L t - \st{R}^* (\gluino{a})^T T^a \cc P_R t
  \right)\nonumber \\
  & +\left(-y_t^2+\frac{g_3^2}{2}\right)(\st{L}^*\st{R})(\st{L}\st{R}^*)-\frac{g_3^2}{6}|\st{L}|^2|\st{R}|^2,
\end{align}
where we use the same notation as in \secref{sec: gluinoOut} with $t$
being the top quark, defined as a Dirac fermion, and
$\gluino{a} = \ccfield{(\gluino{a})}$ denotes the gluino, which is a
Majorana fermion.  The complex scalar fields $\st{L}$ and $\st{R}$
represent the stops.
In the following we determine the one-loop Wilson coefficients of the
following operators in the \EFT:
\begin{align}
\Lag_\EFT^{1\Loop} & \supset c_{t_L} \bar{t}_Li\slashed{\partial} t_L+c_{t_R} \bar{t}_Ri\slashed{\partial} t_R+c_{\st{L}} \partial_\mu\st{L}^*\partial^\mu \st{L}-\delta m_{\sq} ^2 |\st{L}|^2+c_{\st{R}} \partial_\mu\st{R}^*\partial^\mu \st{R}-\delta m_{\su} ^2 |\st{R}|^2 
\nonumber \\ & \quad +c^L_{41} \left(\st{Li}^*  \st{Li} \right)^2+c^L_{42} \left(\st{Li}^*  \st{Lj} \right) \left(\st{Lj}^*  \st{Li}\right)+c^R_{4} \left(\st{R}^* \st{R}\right)^2\nonumber 
\\ & \quad +c^{LR}_{41} \left(\st{Li}^* \st{Li}\right)\left(\st{Rj}^* \st{Rj}\right)+c^{LR}_{42} \left(\st{Li}^* \st{Lj}\right)\left(\st{Rj}^* \st{Ri}\right)+ c_G G^a_{\mu \nu} G_a ^{\mu \nu}\nonumber 
\\ & \quad +[c^{LL} _{51} (\bar{t}_{Li} T^a _{ij} \st{Lj})(\ccfield{t}_{Rk} T^a _{kl} \st{Ll})+c^{LL} _{52} (\st{Li}^* T^a _{ij} \overline{\ccfield{t_{Rj}}})(\st{Lk}^* T^a _{kl} t_{Ll})+(L \leftrightarrow R)]\nonumber 
\\ & \quad +[c^{LR} _{51} (\bar{t}_{Li} T^a _{ij} \st{Lj})(\st{Rk}^* T^a_{kl} t_{Rl})+c^{LR} _{52} (\st{Li} \st{Ri}^*) (\bar{t}_{Lj} t_{Rj})+(L \leftrightarrow R)]
\nonumber \\ & \quad +c_{61} ^L (\tilde{t}_{Li}^* \tilde{t}_{Li})^3+c_{62} ^L (\tilde{t}_{Li}^* \tilde{t}_{Li})(\tilde{t}_{Lj}^* \tilde{t}_{Lk})(\tilde{t}_{Lk}^* \tilde{t}_{Lj})+c_{63} ^L (\tilde{t}_{Li}^* \tilde{t}_{Lj})(\tilde{t}_{Lj}^* \tilde{t}_{Lk})(\tilde{t}_{Lk}^* \tilde{t}_{Li})+c_6^R (\tilde{t}_{Ri}^* \tilde{t}_{Ri})^3
\nonumber \\ & \quad +[c_{61} ^{LR} (\tilde{t}_{Li}^* \tilde{t}_{Li})^2(\tilde{t}_{Ri}^* \tilde{t}_{Ri}) +c_{62} ^{LR} (\tilde{t}_{Li}^* \tilde{t}_{Li})(\tilde{t}_{Lj}^* \tilde{t}_{Lk})(\tilde{t}_{Rk}^* \tilde{t}_{Rj})+c_{63} ^{LR} (\tilde{t}_{Li}^* \tilde{t}_{Lj})(\tilde{t}_{Lj}^* \tilde{t}_{Li})(\tilde{t}_{Rk}^* \tilde{t}_{Rk}) \nonumber \\ & \quad + c_{64} ^{LR} (\tilde{t}_{Li}^* \tilde{t}_{Lj})(\tilde{t}_{Lj}^* \tilde{t}_{Lk})(\tilde{t}_{Rk}^* \tilde{t}_{Ri})+c_{61} ^{RL} (\tilde{t}_{Ri}^* \tilde{t}_{Ri})^2(\tilde{t}_{Li}^* \tilde{t}_{Li}) +c_{62} ^{RL} (\tilde{t}_{Ri}^* \tilde{t}_{Ri})(\tilde{t}_{Rj}^* \tilde{t}_{Rk})(\tilde{t}_{Lk}^* \tilde{t}_{Lj})]
\nonumber \\ & \quad +[c_{61} ^{L^\mu L_\mu}\left(\bar{t}_{Li} \gamma^\mu t_{Li}\right)\left(\bar{t}_{Lj} \gamma_\mu t_{Lj}\right)+c_{62} ^{L^\mu L_\mu}\left(\bar{t}_{Li} \gamma^\mu t_{Lj}\right)\left(\bar{t}_{Lj} \gamma_\mu t_{Li}\right)+(L \leftrightarrow R)]\nonumber 
 \\ & \quad +c_{61}^{(LR)^\mu (RL)_\mu} \left(\overline{\ccfield{t_{Ri}}} \gamma^\mu t_{Rj}\right)\left(\bar{t}_{Rj} \gamma_\mu \ccfield{t}_{Ri}\right)+ c_{62}^{(LR)^\mu (RL)_\mu} \left(\overline{\ccfield{t_{Rj}}} \gamma^\mu t_{Ri}\right)\left(\bar{t}_{Rj} \gamma_\mu \ccfield{t}_{Ri}\right)\nonumber \\ 
 & \quad + [c_{61} ^{LL}\left(\overline{\ccfield{t_{Ri}}} t_{Li}\right)\left(\bar{t}_{Lj} \ccfield{t}_{Rj}\right)+c_{62}^{LL}\left(\overline{\ccfield{t_{Ri}}} t_{Lj}\right)\left(\bar{t}_{Lj} \ccfield{t}_{Ri}\right)+(L\leftrightarrow R)]\nonumber 
 \\ & \quad +c_{61} ^{(LR)(RL)} \left(\bar{t}_{Ri} t_{Lj}\right)\left(\bar{t}_{Lj} t_{Ri}\right) +c_{62} ^{(LR)(RL)} \left(\bar{t}_{Rj}t_{Li}\right)\left(\bar{t}_{Lj} t_{Ri}\right).
 \label{eq: gluonOutFirstEFTLag}
\end{align}
These operators represent all derived one-loop stop interactions in
the gaugeless limit and in the unbroken phase,
without contributions from higher-dimensional operators with covariant derivatives.  Terms which
involve SUSY particles beyond the stop are omitted for brevity.  In
\eqref{eq: gluonOutFirstEFTLag} the color indices $i,j,k=1,2,3$ and
$a=1,\ldots,8$ are written out explicitly.  Note that in general
$\Lag_\EFT^{1\Loop}$ contains $SU(2)_L$ and $SU(3)_C$ invariant terms of the
form
$(\tilde{q}^\dagger_{Li}\tilde{q}_{Li})(\tilde{q}^\dagger_{Lj}\tilde{q}_{Lj})$
and
$(\tilde{q}^\dagger_{Li}\tilde{q}_{Lj})(\tilde{q}^\dagger_{Lj}\tilde{q}_{Li})$,
where the $SU(2)_L$ indices are contracted within parentheses, but the
color indices are contracted differently among the terms.  In
\eqref{eq: gluonOutFirstEFTLag}, however, the corresponding terms with
the couplings $c_{41}^L$ and $c_{42}^L$ have the same structure,
because we have omitted the sbottom quark.

The dimension 5 operators have contributions already at tree-level,
which stem from the insertion of the gluino background field
$\gluinocl$ into the Lagrangian of the MSSM.  The necessary part of the
gluino background field can be extracted from the equation of motion
\begin{align}
[\cc (i\slashed{\partial}-m_{\gluino{}})]_{\alpha \beta} (\gluinocl)_\beta^a=\sqrt{2} g_3 \left(-
    \bar{t}_{L \alpha} T^a \st{L} + \bar{t}_{R \alpha}  T^a \st{R}
    + \st{L}^* T^a (\cc t_L)_\alpha - \st{R}^* T^a (\cc t_R)_\alpha
  \right),
\end{align}
which yields
\begin{align}
(\gluinocl)_\beta^a &= \sqrt{2} g_3 (i\slashed{\partial}-m_{\gluino{}})_{\beta \alpha}^{-1} \left[-
    (\bar{t}_L \cc) _\alpha T^a \st{L} + (\bar{t}_R \cc)_\alpha  T^a \st{R}
    + \st{L}^* T^a  t_{L\alpha} - \st{R}^* T^a t_{R\alpha} \right]
    \\ &= \frac{\sqrt{2} g_3}{m_{\gluino{}}} \left[
    (\bar{t}_L \cc) _\beta T^a \st{L} - (\bar{t}_R \cc)_\beta  T^a \st{R}
    - \st{L}^* T^a  t_{L\beta} + \st{R}^* T^a t_{R\beta} + \cdots \right] ,
    \label{eq: ClassicalGluinoField}
\end{align}
where the ellipsis designate higher order terms of
$\order{\partial/m_{\gluino{}}}$ with at least one
derivative. Inserting \eqref{eq: ClassicalGluinoField} into both the
kinetic term of the gluino and the interaction Lagrangian one finds
the tree-level values of $c_{5i} ^{AB}$ ($A,B\in \{L,R\}$) to be
\begin{align}
c_{51} ^{LL,\tree}&=c_{52} ^{LL,\tree}=c_{51} ^{RR,\tree}=c_{52} ^{RR,\tree}=\frac{g^2_3}{m_{\gluino{}}}, \\
c_{51} ^{LR,\tree}&=c_{51} ^{RL,\tree}=-\frac{2g^2_3}{m_{\gluino{}}}, \\
c_{52} ^{LR,\tree}&=c_{52} ^{RL,\tree}=0.
\end{align}
At one-loop the relevant contributions from the UOLEA are
\begin{align}
\frac{1}{\kappa}\Lag_\EFT^{1\Loop} = \tr  \Big\{&(-\ZI [q^4]^{31} _{\gluino{} 0} +\frac{m^2_{\gluino{}}}{12} \ZI [q^2] ^{22} _{\gluino{} 0}) \gamma_\mu [P^\nu,(\mathbf{X}_{\Xi \xi})^a_i] \gamma ^\mu [P_\nu,(\mathbf{X}_{\xi \Xi})^a_i] \nonumber \\
&  +(-2\ZI [q^4]^{31} _{\gluino{} 0} +\frac{m^2_{\gluino{}}}{6} \ZI [q^2] ^{22} _{\gluino{} 0}) \gamma_\mu [P^\mu,(\mathbf{X}_{\Xi \xi})^a_i] \gamma ^\nu [P_\nu,(\mathbf{X}_{\xi \Xi})^a_i] \nonumber \\ &  + (-\ZI [q^{2}]^{12} _{\gluino{}0}-2 m^2_{\phi_i} \ZI[q^2] ^{13} _{\gluino{}0}) (\mathbf{X}_{\phi \Xi})_i \gamma ^\mu [P_\mu,(\mathbf{X}_{\Xi \phi})_i]\nonumber
\\
&  +\frac{1}{4} \ZI [q^2] ^{22}_{\gluino{}0} (\mathbf{X}_{\phi \Xi})_i \gamma^\mu (\mathbf{X}_{\Xi \phi})_j (\mathbf{X}_{\phi \Xi})_j \gamma_\mu (\mathbf{X}_{\Xi \phi})_i
\nonumber \\ &  -\frac{1}{2}m_{\gluino{}} \ZI ^{12}_{\gluino{}0}(\mathbf{X}_{\phi \phi})_{ij} (\mathbf{X}_{\phi \Xi})_j  (\mathbf{X}_{\Xi \phi})_i 
\nonumber \\ &  +\frac{1}{4} m^2_{\gluino{}} \ZI  ^{22}_{\gluino{}0} (\mathbf{X}_{\phi \Xi})_i (\mathbf{X}_{\Xi \phi})_j (\mathbf{X}_{\phi \Xi})_j (\mathbf{X}_{\Xi \phi})_i-\frac{1}{2} \ZI [q^2]^{11} _{{\gluino{}} 0} \gamma^\mu (\mathbf{X}_{\Xi \xi})_i \gamma_\mu (\mathbf{X}_{\xi \Xi})_i \nonumber 
\\ &  -\frac{1}{4}m^2_{\gluino{}} \ZI [q^2] ^{22} _{\gluino{}0} (\mathbf{X}_{\Xi \xi})^a_i \gamma^\mu (\mathbf{X}_{\xi \Xi})^b_i (\mathbf{X}_{\Xi \xi})^b_j \gamma_\mu (\mathbf{X}_{\xi \Xi})^a_j 
\nonumber 
\\
&  -\frac{1}{4} \ZI [q^4] ^{22} _{\gluino{}0} g_{\mu \nu \rho \sigma}  (\mathbf{X}_{\Xi \xi})^a_i \gamma^\mu (\mathbf{X}_{\xi \Xi})^b_i \gamma^\nu (\mathbf{X}_{\Xi \xi})^b_j \gamma^\rho (\mathbf{X}_{\xi \Xi})^a_j \gamma^\sigma
\nonumber
\\
&  -\frac{1}{2}m^2 _{\gluino{}} \ZI [q^4] ^{33} _{\gluino{}0}  g_{\mu \nu \rho \sigma}  (\mathbf{X}_{\Xi \xi})^a_i \gamma^\mu (\mathbf{X}_{\xi \Xi})^b_i  (\mathbf{X}_{\Xi \xi})^b_j \gamma^\nu (\mathbf{X}_{\xi \Xi})^c_j  \gamma^\rho (\mathbf{X}_{\Xi \xi})^c_k \gamma^\sigma (\mathbf{X}_{\xi \Xi})^a_k
\nonumber
\\
&  -\frac{1}{6} \ZI [q^6] ^{33} _{\gluino{}0} g_{\mu \nu \rho \sigma \kappa \lambda} (\mathbf{X}_{\Xi \xi})^a_i \gamma^\mu (\mathbf{X}_{\xi \Xi})^b_i \gamma^\nu (\mathbf{X}_{\Xi \xi})^b_j \gamma^\rho (\mathbf{X}_{\xi \Xi})^c_j  \gamma^\sigma (\mathbf{X}_{\Xi \xi})^c_k \gamma^\kappa (\mathbf{X}_{\xi \Xi})^a_k \gamma^\lambda
\nonumber \\ &  +\frac{1}{6}\ZI ^{2} _{\gluino{}}[P_\mu,P_\nu][P^\mu,P^\nu]\Big\},
\label{eq: UOLEAContrGluino}
\end{align}
where $g_{\mu \nu \cdots}$ is the combination of metric tensors which
is totally symmetric in all indices, see
\appref{sec:loop_functions}. The derivatives with respect to the stops
and the gluino have already been calculated in
\secref{sec:matching_MSSM_to_SMEFT} and are given by
\begin{align}
  \mathbf{X}_{\phi \Xi} &=
  \begin{pmatrix}
    X_{\sigma ^* \Lambda} \\
    X_{\sigma \Lambda}
  \end{pmatrix}
  =
  \begin{pmatrix}
    (X_{\st{L}^* \gluino{a}})_{i\alpha}^a \\
    (X_{\st{R}^* \gluino{a}})_{i\alpha}^a \\
    (X_{\st{L} \gluino{a}})_{i\alpha}^a \\
    (X_{\st{R} \gluino{a}})_{i\alpha}^a
  \end{pmatrix}
  = \sqrt{2} g_3
  \begin{pmatrix}
    T^a_{ij} (\cc P_L t_j)_{\alpha} \\
     -T^a_{ij} (\cc P_R t_j)_{\alpha} \\
     -(\bar{t}_j P_R)_{\alpha} T^a_{ji} \\
    (\bar{t}_j P_L)_{\alpha} T^a_{ji}
  \end{pmatrix},
  \\
  \mathbf{X}_{\Xi \phi} &=
  \begin{pmatrix}
    \cc^{-1} X_{\Lambda \sigma}, & \cc^{-1} X_{\Lambda \sigma ^*}
  \end{pmatrix} \\
  &= (\cc^{-1})_{\alpha\beta}
  \begin{pmatrix}
    (X_{\gluino{a} \st{L}})_{i\beta}^a, &
    (X_{\gluino{a} \st{R}})_{i\beta}^a, &
    (X_{\gluino{a} \st{L}^*})_{i\beta}^a, &
    (X_{\gluino{a} \st{R}^*})_{i\beta}^a
  \end{pmatrix} \\
  &= \sqrt{2} g_3
  \begin{pmatrix}
    -(\bar{t}_j P_R \cc)_{\alpha} T^a_{ji}, &
   (\bar{t}_j P_L \cc)_{\alpha} T^a_{ji}, &
   T^a_{ij} (P_L t_j)_{\alpha}, &
    -T^a_{ij} (P_R t_j)_{\alpha}
  \end{pmatrix},
\end{align}
the difference being that the stops are now considered to be light
fields.
For the purpose of this application we also need the derivatives with
respect to a top and a gluino, which read
\begin{align}
(X_{\bar{t} \gluino{a}})_{i \alpha \beta}^a&=-\sqrt{2}g_3T^a_{ij}\left[(P_R)_{\alpha\beta}\st{Lj}-(P_L)_{\alpha\beta}\st{Rj}\right],\\
(X_{t \gluino{a}})_{i \alpha \beta}^a&=-\sqrt{2}g_3T^a_{ji}\left[-\st{Lj}^*(\cc P_L)_{\beta \alpha}+\st{Rj}^*(\cc P_R)_{\beta \alpha}\right],\\
(X_{\gluino{a}\bar{t}})_{i \alpha \beta}^a&=\sqrt{2}g_3T^a_{ij}\left[(P_R)_{\beta \alpha}\st{Lj}-(P_L)_{\beta \alpha}\st{Rj}\right],\\
(X_{ \gluino{a} t})_{i \alpha \beta}^a&=\sqrt{2}g_3T^a_{ji}\left[-\st{Lj}^*(\cc P_L)_{ \alpha \beta}+\st{Rj}^*(\cc P_R)_{\alpha \beta}\right],
\end{align}
and are collected into
\begin{align}
\mathbf{X}_{\Xi \xi}&=\begin{pmatrix}
\cc ^{-1} X_{\Lambda \omega}, & \cc ^{-1} X_{\Lambda \bar{\omega}} \cc ^{-1}
\end{pmatrix}\\
&=\begin{pmatrix}
(\cc ^{-1}X_{ \gluino{a} t})_{i \alpha \beta}^a , & (\cc ^{-1} X_{\gluino{a}\bar{t}} \cc ^{-1})_{i \alpha \beta}^a 
\end{pmatrix} \\
&= \begin{pmatrix}
-\sqrt{2}g_3T^a_{ji}\left[\st{Lj}^*(P_L)_{ \alpha \beta}-\st{Rj}^*(P_R)_{ \alpha \beta}\right] , & -\sqrt{2}g_3T^a_{ij}\left[(P_R)_{\alpha \beta}\st{Lj}-(P_L)_{\alpha \beta}\st{Rj}\right]
\end{pmatrix}, \\ 
\mathbf{X}_{\xi \Xi}&=\begin{pmatrix}
 X_{\bar{\omega} \Lambda} \\
  \cc^{-1} X_{\omega \Lambda} 
\end{pmatrix}=\begin{pmatrix}
 (X_{\bar{t} \gluino{a}})_{i \alpha \beta}^a \\
  (\cc^{-1}X_{t \gluino{a}})_{i \alpha \beta}^a
\end{pmatrix}=\begin{pmatrix}
 -\sqrt{2}g_3T^a_{ij}\left[(P_R)_{\alpha\beta}\st{Lj}-(P_L)_{\alpha\beta}\st{Rj}\right]\\
  -\sqrt{2}g_3T^a_{ji}\left[\st{Lj}^*(P_L)_{\alpha \beta}-\st{Rj}^*(P_R)_{\alpha \beta}\right]
\end{pmatrix}.
\end{align}
Finally we give the derivatives with respect to two stops
\begin{align}
\mathbf{X}_{\phi \phi} &= \begin{pmatrix}
\mathbf{Y}_{\phi \phi} & \mathbf{0}_{2\times2} \\
\mathbf{0}_{2\times2} & (\mathbf{Y}_{\phi \phi})^*
\end{pmatrix},\\
\mathbf{Y}_{\phi \phi} &=
\begin{pmatrix}
x_t \st{Rj}^* \st{Ri}-\frac{g_3 ^2}{6 }\st{R}^* \st{R} \delta_{ij} && x_t  \delta_{ij} \st{L} \st{R}^*-\frac{g_3^2}{6}\st{Li} \st{Rj}^* \\
x_t  \delta_{ij} \st{L}^* \st{R}-\frac{g_3^2}{6}\st{Ri} \st{Lj}^* && x_t \st{Lj}^* \st{Li}-\frac{g_3 ^2}{6 }\st{L}^* \st{L} \delta_{ij}
\end{pmatrix},
\end{align}
where we have introduced the abbreviation $x_t \equiv y_t^2-g_3 ^2/2$.
Substituting these derivatives into \eqref{eq: UOLEAContrGluino} and
summing over all indices one finds
\begin{align}
c_{t_L}&=\frac{16}{3}g^2_3\left(\ZI [q^{2}]^{12} _{\gluino{}0}+2 m^2_{\tilde{q}} \ZI[q^2] ^{13} _{\gluino{}0}\right), \\
c_{t_R}&=\frac{16}{3}g^2_3\left(\ZI [q^{2}]^{12} _{\gluino{}0}+2 m^2_{\tilde{u}} \ZI[q^2] ^{13} _{\gluino{}0}\right), \\
c_{\st{L}}&=c_{\st{R}}=\frac{32}{3}g^2_3(d+2)\left(-\ZI [q^{4}]^{31} _{\gluino{}0}+ \frac{m^2_{\tilde{q}}}{2} \ZI[q^2] ^{22} _{\gluino{}0}\right),\\
c_{61}^{L^\mu L_\mu}&=c_{61} ^{R^\mu R_\mu}=\frac{7}{6}g^4_3 \ZI [q^2] ^{22}_{\gluino{}0}, \\
c_{62} ^{L^\mu L_\mu}&=c_{62} ^{ R^\mu R_\mu}=\frac{1}{18}g^4_3 \ZI [q^2] ^{22}_{\gluino{}0}, \\
c_{61} ^{(LR)^\mu (RL)_\mu}&=\frac{10}{9}g_3^4 \ZI [q^2] ^{22}_{\gluino{}0}, \\
c_{62}^{(LR)^\mu (RL)_\mu}&=-\frac{2}{9}g_3^4 \ZI [q^2] ^{22}_{\gluino{}0}, \\
c_{61} ^{LL}&=c_{61} ^{ R R}=\frac{5}{18}g^4_3 m^2_{\gluino{}}\ZI [q^2] ^{22}_{\gluino{}0}, \\
c_{62} ^{LL}&=c_{62} ^{ R R}=-\frac{1}{6}g^4_3 m^2_{\gluino{}}\ZI [q^2] ^{22}_{\gluino{}0}, \\
c_{61} ^{(LR)(RL)}&=\frac{7}{6}g_3^4 m^2_{\gluino{}}\ZI [q^2] ^{22}_{\gluino{}0}, \\
c_{62} ^{(LR) (RL)}&=\frac{1}{18}g_3^4 m^2_{\gluino{}}\ZI [q^2] ^{22}_{\gluino{}0}, \\
\delta m_{\sq} ^2 &= \delta m_{\su} ^2 =\frac{16}{3}dg^2 _3 \ZI[q^2] ^{11} _{\gluino{}0}, \\
c^L _{41} &= -\frac{40}{9}m^2_{\gluino{}}g_3^4 \ZI[q^2] ^{22} _{\gluino{}0}-\frac{1}{9}d(d+2)g_3^4 \ZI[q^4] ^{22} _{\gluino{}0}, \\
c^R _{4} &= -\frac{16}{3}m^2_{\gluino{}}g_3^4 \ZI[q^2] ^{22} _{\gluino{}0}-\frac{22}{9}d(d+2)g_3^4 \ZI[q^4] ^{22} _{\gluino{}0}, \\
c^L _{42} &= \frac{8}{3}m^2_{\gluino{}}g_3^4 \ZI[q^2] ^{22} _{\gluino{}0}-\frac{7}{3}d(d+2)g_3^4 \ZI[q^4] ^{22} _{\gluino{}0}, \\
c^{LR} _{41} &= -\frac{8}{9}m^2_{\gluino{}}g_3^4 \ZI[q^2] ^{22} _{\gluino{}0}-\frac{20}{9}d(d+2)g_3^4 \ZI[q^4] ^{22} _{\gluino{}0}, \\
c^{LR} _{42} &= -\frac{56}{3}m^2_{\gluino{}}g_3^4 \ZI[q^2] ^{22} _{\gluino{}0}+\frac{4}{9}d(d+2)g_3^4 \ZI[q^4] ^{22} _{\gluino{}0}, \\
c^{L} _{61} &= \frac{1}{54}d(d+2)g_3^6 m^2_{\gluino{}} \ZI[q^4] ^{33} _{\gluino{}0}+\frac{2}{81}d(d^2+6d+8)g_3^6  \ZI[q^6] ^{33} _{\gluino{}0},	 \\
c^{L} _{62} &=- \frac{2}{3}d(d+2)g_3^6 m^2_{\gluino{}} \ZI[q^4] ^{33} _{\gluino{}0}-\frac{2}{9}d(d^2+6d+8)g_3^6  \ZI[q^6] ^{33} _{\gluino{}0}, \\
c^{L} _{63} &= \frac{1}{2}d(d+2)g_3^6 m^2_{\gluino{}} \ZI[q^4] ^{33} _{\gluino{}0}-\frac{4}{3}d(d^2+6d+8)g_3^6  \ZI[q^6] ^{33} _{\gluino{}0}, \\
c^{R} _{6} &=-\frac{4}{27}d(d+2)g_3^6 m^2_{\gluino{}} \ZI[q^4] ^{33} _{\gluino{}0}-\frac{124}{81}d(d^2+6d+8)g_3^6  \ZI[q^6] ^{33} _{\gluino{}0}, \\
c^{LR} _{61} &= \frac{1}{18}d(d+2)g_3^6 m^2_{\gluino{}} \ZI[q^4] ^{33} _{\gluino{}0} +\frac{2}{27}d(d^2+6d+8)g_3^6  \ZI[q^6] ^{33} _{\gluino{}0},\\
c^{LR} _{62} &=- \frac{12}{9}d(d+2)g_3^6 m^2_{\gluino{}} \ZI[q^4] ^{33} _{\gluino{}0}-\frac{10}{9}d(d^2+6d+8)g_3^6  \ZI[q^6] ^{33} _{\gluino{}0}, \\
c^{LR} _{63} &= -\frac{1}{6}d(d+2)g_3^6 m^2_{\gluino{}} \ZI[q^4] ^{33} _{\gluino{}0}-\frac{14}{9}d(d^2+6d+8)g_3^6  \ZI[q^6] ^{33} _{\gluino{}0}, \\
c^{LR} _{64} &= \frac{2}{9}d(d^2+6d+8)g_3^6  \ZI[q^6] ^{33} _{\gluino{}0}, \\
c^{RL} _{61} &= -\frac{1}{9}d(d+2)g_3^6 m^2_{\gluino{}} \ZI[q^4] ^{33} _{\gluino{}0}-\frac{40}{27}d(d^2+6d+8)g_3^6  \ZI[q^6] ^{33} _{\gluino{}0},\\
c^{RL} _{62} &=- \frac{12}{9}d(d+2)g_3^6 m^2_{\gluino{}} \ZI[q^4] ^{33} _{\gluino{}0}+\frac{8}{9}d(d^2+6d+8)g_3^6  \ZI[q^6] ^{33} _{\gluino{}0}, \\
c_{51} ^{LR\text{,1\Loop}}&=c_{51} ^{RL\text{,1\Loop}}=-\frac{g_3^4}{3}m_{\gluino{}} \ZI ^{12}_{\gluino{}0}, \\
c_{52} ^{LR\text{,1\Loop}}&=c_{52} ^{RL\text{,1\Loop}}=-\frac{8}{3}g_3^4 x_t m_{\gluino{}} \ZI ^{12}_{\gluino{}0}, \\
c_G&=-\frac{g_3^2}{2}\ZI ^{2} _{\gluino{}}.
\end{align}
In the calculation of these corrections the relations
$g^{\mu\nu}g_{\mu\nu} = d = 4 - \eps$ and \eqref{eq:genrel} were used
repeatedly.  The one-loop corrections $\delta m_{\sq}^2$ and
$\delta m_{\su}^2$ to the third generation squark mass parameters have
already been calculated in \cite{Aebischer:2017aqa} and our results
agree with the expressions found there.

Since supersymmetry is only softly broken in the MSSM it is convenient
to use DRED as a regulator. Once the gluino is
integrated out from the theory, supersymmetry is explicitly broken and
it is natural to regularize the EFT in
DREG\@.  This switch in the regularization scheme introduces further
contributions to the couplings of the EFT coming from the epsilon
scalars.  In the formalism of the UOLEA the relevant operators which
contribute here are given by \cite{Summ:2018oko}
\begin{align}
\begin{split}
  \frac{\eps}{\kappa} \Lag^{1\ell}_\text{reg} =
&-\sum _{i} (m^2_{\eps})_{i} (\epsdim{X}^\mu _{\eps \eps \mu})_{ii}
 + \frac{1}{2} \sum_{ij} (\epsdim{X}^{\mu}_{\eps \eps \nu})_{ij} (\epsdim{X}^{\nu}_{\eps \eps \mu})_{ji}  \\ 
&+\sum_{ij} 2^{c_{F_j}} \left\{2 m_{\psi j} (\epsdim{X}^\mu_{\eps \psi})_{ij} (\epsdim{X} _{\bar{\psi} \eps \mu})_{ji} + (\epsdim{X}^\mu_{\eps \psi})_{ij} \gamma^\nu \left[P_\nu,(\epsdim{X}_{\bar{\psi} \eps \mu})_{ji}\right]\right\} \\
&-\sum_{i j k} 2^{c_{F_j}+c_{F_k}-1} (\epsdim{X}^\mu_{\eps \psi})_{ij} \gamma ^\nu (X_{\bar{\psi} \psi})_{jk} \gamma_{\nu} (\epsdim{X}_{\bar{\psi} \eps \mu})_{ki} \\ 
& + \frac{\eps}{12} \tr\left[ G'_{\mu \nu} G'^{\mu \nu} \right],
\end{split}
\label{eq:epsilon-scalar contributions}
\end{align}
The $\epsdim{X}$ operators are projections of the corresponding
$4$-dimensional ones $\fourdim{X}$ onto the $\eps$-dimensional
$Q\eps S$ space, i.e.
\begin{align}
  \epsdim{X}^\mu &= \epsdim{g}^\mu_\sigma \fourdim{X}^\sigma, \\
  \epsdim{X}^{\mu\nu} &= \epsdim{g}^\mu_\sigma \epsdim{g}^\nu_\rho \fourdim{X}^{\sigma\rho},
\end{align}
see \appref{sec:DREG_DRED}.  Furthermore,
$G'_{\mu\nu} = -ig_3 G^a_{\mu\nu} T^a$ is the gluon field strength
tensor.  For the top quark (a Dirac fermion) we have $c_F = 0$, and
for the gluino (a Majorana fermion) $c_F = 1$.  From
\eqref{eq:epsilon-scalar contributions} we obtain the following
additional contributions to the couplings of the EFT
\begin{align}
(\delta m^2 _{\sq})_\eps &=(\delta m ^2 _{\su})_\eps=-\frac{4}{3} g_3^2 m_\eps ^2, \label{eq:delta_m2_eps} \\
(c_{t_L})_\eps &= (c_{t_R})_\eps=\frac{4}{3}g_3^2, \\
(c^L_{41})_\eps &= \frac{1}{72}g_3^4, \\
(c^L_{42})_\eps&=\frac{7}{24}g_3^4, \\
(c^R_{4})_\eps &= \frac{11}{36}g_3^4, \\
(c^{LR}_{41})_\eps &= \frac{1}{36}g_3^4, \\
(c^{LR}_{42})_\eps &= \frac{7}{12}g_3^4, \\
(c^{LL}_{51})_\eps &= (c^{LL}_{52})_\eps = (c^{RR}_{51})_\eps=(c^{RR}_{52})_\eps = \frac{3g_3^4 }{2 m_{\gluino{}}}d, \\
(c^{LR}_{51})_\eps &= (c^{RL}_{52})_\eps = -\frac{3g_3^4 }{m_{\gluino{}}}d, \\
(c_G)_\eps &= -\frac{g_3^2}{4}.
\end{align}
%One may write the effective Lagrangian in terms of the effective gauge coupling $\hat{g}_3=g_3+\delta g_3$ and the re-scaled fields $\hat{t}_L=(1+Z_L/2)t_L$, $\hat{t}_R=(1+Z_R/2)t_R$ and $\hat{g}^a_\mu=(1+Z_g/2)g^a_\mu$ as
%\begin{align}
%\Lag_\EFT^{1\Loop}&=\bar{\hat{t}}_L i\slashed{\partial}\hat{t}_L+\bar{\hat{t}}_R i\slashed{\partial}\hat{t}_R-\frac{1}{4}G^{\mu \nu}_aG^a_{\mu \nu}+\cdots \nonumber \\ 
%&= \bar{t}_L i\slashed{\partial}t_L+Z_L \bar{t}_L i\slashed{\partial}t_L+\bar{t}_R i\slashed{\partial}t_R+Z_R \bar{t}_R i\slashed{\partial}t_R-(1+Z_g)\frac{1}{4}G^{\mu \nu}_aG^a_{\mu \nu}\nonumber \\ & \quad-\frac{1}{2}\left(\delta g_3+\frac{1}{2}g_3 Z_g\right)G_{\mu \nu}^a f^{abc}g^\mu _b g^\nu _c.
%\label{eq: gluonOutFinalEFTLag}
%\end{align}
%Since the last term is not gauge invariant we must have $\delta g_3=-
%\frac{1}{2} g_3 Z_g$, which is a well-known Slavnov-Taylor identity. The remaining terms fix the field re-definitions. Comparing \eqref{eq: gluonOutFinalEFTLag} to \eqref{eq: gluonOutFirstEFTLag} we obtain 
%\begin{align}
%Z_L&=c_{t_L}, \\
%Z_R&=c_{t_R}, \\
%Z_g &= -4c_G, \\
%\delta g_3 &= 2g_3 c_G.
%\end{align}
%
The term $\propto m_\eps^2$ on the r.h.s.\ of
\eqref{eq:delta_m2_eps} can be removed by switching from the \DRbar\ to
the \DRbarPrime\ scheme \cite{Jack:1994rk}, which involves shifting
$m^2_{\sq}$ and $m ^2_{\su}$ by finite terms.
Notice also that the one-loop DRED--DREG conversion corrections to the
coefficients of the dimension 5 operators arise from the third line of
\eqref{eq:epsilon-scalar contributions}, which among other terms
contains the term
\begin{align}
  (\epsdim{X}^\mu_{\eps t})\gamma ^\nu (X_{\bar{t} \gluino{}}) \gamma_{\nu} (\epsdim{X}_{\bar{\gluino{}} \eps \mu}).
\end{align}
Here $(\epsdim{X}_{\bar{\gluino{}} \eps \mu})$ has an explicit
dependence on the gluino spinor $\gluino{}$,
\begin{align}
(\epsdim{X}_{\bar{\gluino{}} \eps \mu})^{ba}=\frac{ig_3}{2}\epsdim{\gamma}^\mu f^{abc}\gluino{c},
\end{align}
which must be eliminated by inserting the background field from
\eqref{eq: ClassicalGluinoField}. As noted above the threshold
corrections for the two stop masses agree with the results derived in
\cite{Aebischer:2017aqa} when the effect of the sbottom quarks is
neglected.

%%%%%%%%%%%%%%%%%%%%%%%%%%%%%%%%%%%%%%%%%%%%%%%%%%

\section{Conclusions}\label{sec:conclusions}

In this paper we have presented an extension of the Universal One-Loop
Effective Action (UOLEA) by all one-loop operators up to dimension 6
for generic theories with scalar and fermionic fields, excluding
operators stemming from open covariant derivatives in the UV
Lagrangian.  Our generic results can be used to derive the analytic
expressions of all one-loop Wilson coefficients up to dimension 6 of
an effective Lagrangian from a given UV theory with heavy scalar or
fermionic particles, as long as second derivatives of the UV
Lagrangian w.r.t.\ the fields do not contain covariant derivatives.
Thus, our new results allow for an application of the UOLEA to a
broader class of UV models than before.

To illustrate and  test our generic results we have applied the UOLEA
to different EFTs of the SM and the MSSM, where parts of the spectrum are heavy.
We were able to reproduce known results from the
literature, including the prediction of some one-loop Wilson
coefficients of higher-dimensional operators of the SMEFT.

We have published our results in form of the two ancillary Mathematica
files \texttt{UOLEA.m} and \texttt{LoopFunctions.m}, which allow for a
direct use of our expressions and a potential implementation into
generic tools such as \CoDEx or spectrum generator generators such as
\SARAH\ and \FS.

%%%%%%%%%%%%%%%%%%%%%%%%%%%%%%%%%%%%%%%%%%%%%%%%%%

\acknowledgments{%
We kindly thank Jérémie Quevillon for helpful discussions regarding
the UOLEA. BS would like to thank the Institute for Theoretical Physics in Heidelberg, where part of this work was completed, for its hospitality. This research was supported by the German DFG
Collaborative Research Centre \textit{P\textsuperscript{3}\!H: Particle Physics Phenomenology after the Higgs Discovery}
(CRC TRR 257).
}

%%%%%%%%%%%%%%%%%%%%%%%%%%%%%%%%%%%%%%%%%%%%%%%%%%

\appendix

\section{Fermionic shifts}

\label{sec:shifts}
In this section we discuss the consistency of the shift
\eqref{eq:xishift}. The treatment of the shift given in
\eqref{eq:Xishift} is analogous but somewhat more involved.  Since
$\xi$ is a multiplet of Majorana-like component spinors,
for the shift
\begin{align}
  \delta \xi' = \delta \xi+\mathbf{\Delta}_\xi^{-1}\left[\tilde{\mathbf{X}}_{\xi \Xi} \delta \Xi-\tilde{\mathbf{X}}_{\xi \Phi} \delta \Phi-\tilde{\mathbf{X}}_{\xi \phi} \delta \phi\right]
\label{eq:xishift2}
\end{align}
to be consistent it is necessary and sufficient that
\begin{align}
\left(\mathbf{\Delta}_\xi^{-1}\left[\tilde{\mathbf{X}}_{\xi \Xi} \delta \Xi-\tilde{\mathbf{X}}_{\xi \Phi} \delta \Phi-\tilde{\mathbf{X}}_{\xi \phi} \delta \phi\right]\right)^{T}=\left[\delta \Xi^T \tilde{\mathbf{X}}_{\Xi \xi}+\delta \Phi^T \tilde{\mathbf{X}}_{\Phi \xi}+\delta \phi^T \tilde{\mathbf{X}}_{\phi \xi}\right]\overleftarrow{\mathbf{\Delta}}_\xi^{-1}.
\label{eq:shiftCondition}
\end{align}
In the following we show that \eqref{eq:shiftCondition} holds.
We first construct $\mathbf{\Delta}_\xi ^{-1}$ in position space through
its Neumann series\footnote{In what follows we always write the whole
  series. In practice, however, we are only ever interested in a
  finite number of terms with all higher order terms being suppressed
  by higher powers of couplings.}
\begin{align}
\mathbf{\Delta}_\xi ^{-1}(x,y) &= \sum _{n=0} ^{\infty} \left(\prod _{\substack{i=1 \\ n>0}} ^{n} \int \rd^d x_i \; \mathbf{S}(x_{i-1},x_i) \left(-\mathbf{X}_{\xi \xi}(x_i)\right)\right) \mathbf{S}(x_n,y) \tilde{\mathds{1}} \cc ^{-1} \nonumber \\
& \equiv \sum _{n=0} ^{\infty} \left(\prod _{\substack{i=1 \\ n>0}} ^{n}  \mathbf{S}_{x_{i-1} x_i} \left(-\mathbf{X}_{\xi \xi x_i}\right)\right) \mathbf{S}_{x_n y} \tilde{\mathds{1}} \cc ^{-1}, 
\end{align}
where $x_0\equiv x$ and $\mathbf{S}(x,y)$ is the matrix-valued Green's
function for $(\slashed{P}-M_\xi)$, which itself can be expressed
through a Neumann series. To keep expressions short we also introduced
the convention of denoting space-time points by indices, where repeated
indices are integrated over. We may write
$(\slashed{P}-M_\xi) = (i\slashed{\partial}-M_\xi-\mathbf{A})$ with
\begin{align}
\mathbf{A}=i\sum_j g_j\slashed{A}_j^a T_j^a,
\end{align}
where we sum over all factors of the gauge group for a direct product
group and $T_j^a$ is a block-diagonal matrix which generates the
reducible representation of $\xi$. Due to the fact that $\xi$ contains
$\omega$, $\ccfield{\omega}$ and $\lambda$ (see \eqref{eq:def_xi}),
the generator is of the form
\begin{align}
T^a=\begin{pmatrix}
T^a _{R(\omega)} && 0 && 0 \\
0 && T^a _{\bar{R}(\omega)} && 0 \\
0 && 0 && T^a_{R( \lambda )}
\end{pmatrix},
\end{align}
where $R(\omega)$ is the representation under which $\omega$
transforms, $\bar{R}(\omega)$ its conjugate representation and
$R(\lambda)$ is the representation of $\lambda$, which is necessarily
real. We then have
\begin{align}
\mathbf{S}_{x y}=\sum _{k=0} ^{\infty} \left(\prod _{\substack{i=1 \\ k>0}} ^{k}  \mathbf{S}_{f,x_{i-1} x_i} \mathbf{A}_{x_i}\right) \mathbf{S}_{f, x_k y},
\end{align}
where again $x_0 \equiv x$ and $\mathbf{S}_{f,x y}$ is the matrix
containing the Green's function of the free Dirac equation on its
diagonal. It can be verified by explicit calculation that
\begin{align}
\mathbf{S}_{x y} \left(-i\overleftarrow{\slashed{\partial}_y}-M_\xi-\mathbf{A}_y \right) &= \delta_{x y},
\end{align} 
which means that
\begin{align}
\mathbf{\Delta}_{\xi,x y} ^{-1}\overleftarrow{\mathbf{\Delta}}_{\xi, y}=\delta_{x y}
\end{align}
and therefore
$\overleftarrow{\mathbf{\Delta}}_{\xi,yx}
^{-1}=\mathbf{\Delta}_{\xi,yx}^{-1}$. Hence \eqref{eq:shiftCondition}
reads
\begin{align}
\left(\mathbf{\Delta}_{\xi, x y}^{-1}  \left[\tilde{\mathbf{X}}_{\xi \Xi} \delta \Xi-\tilde{\mathbf{X}}_{\xi \Phi} \delta \Phi-\tilde{\mathbf{X}}_{\xi \phi} \delta \phi\right]_y\right)^{T}=\left[\delta \Xi^T \tilde{\mathbf{X}}_{\Xi \xi}+\delta \Phi^T \tilde{\mathbf{X}}_{\Phi \xi}+\delta \phi^T \tilde{\mathbf{X}}_{\phi \xi}\right]_y\mathbf{\Delta}_{\xi, yx}^{-1}.
\end{align}
It is then useful to calculate
\begin{align}
\cc \tilde{\mathds{1}} \mathbf{S}^T_{xy}&=\cc \tilde{\mathds{1}}\sum _{k=0} ^{\infty} \mathbf{S}^T_{f,x_k y} \left(\prod _{\substack{i=k \\ k>0}} ^{1}  \mathbf{A}^T_{x_i}\mathbf{S}^T_{f,x_{i-1}x_i} \right)  \\
&=  \cc \tilde{\mathds{1}}\sum _{k=0} ^{\infty} \cc \mathbf{S}_{f,y x_k} \cc^{-1} \left(\prod _{\substack{i=k \\ k>0}} ^{1}  \mathbf{A}^T _{x_i}\cc \mathbf{S}_{f,x_i x_{i-1}}\cc^{-1} \right)  \\
&=  -\sum _{k=0} ^{\infty}  \tilde{\mathds{1}} \mathbf{S}_{f,y x_k} \tilde{\mathds{1}} \tilde{\mathds{1}} \left(\prod _{\substack{i=k \\ k>0}} ^{1}   (-\mathbf{A}^t _{x_i})\tilde{\mathds{1}} \tilde{\mathds{1}} \mathbf{S}_{f,x_i x_{i-1}} \tilde{\mathds{1}} \tilde{\mathds{1}} \right)\cc^{-1}
 \\
&=  -\sum _{k=0} ^{\infty}  \mathbf{S}_{f, y x_k} \left(\prod _{\substack{i=k \\ k>0}} ^{1} (-\tilde{\mathds{1}}\mathbf{A}^t_{x_i}\tilde{\mathds{1}})\mathbf{S}_{f,x_i x_{i-1}} \right)\tilde{\mathds{1}}\cc^{-1}  \\
&=  -\sum _{k=0} ^{\infty}  \mathbf{S}_{f,y x_k} \left(\prod _{\substack{i=k \\ k>0}} ^{1}  \mathbf{A}_{x_i} \mathbf{S}_{f,x_i x_{i-1}} \right)\tilde{\mathds{1}}\cc^{-1}  \\
&= -\mathbf{S}_{yx}\tilde{\mathds{1}}\cc ^{-1},
\end{align}
where $\mathbf{A}^t$ means taking the transpose of the gauge group
generators only and we used that
\begin{align}
\tilde{\mathds{1}}\begin{pmatrix}
A && 0 && 0 \\
0 && B && 0 \\
0 && 0 && C 
\end{pmatrix}
\tilde{\mathds{1}}=\begin{pmatrix}
B && 0 && 0 \\
0 && A && 0 \\
0 && 0 && C 
\end{pmatrix}.
\end{align}
We then find
\begin{align}
\left(\mathbf{\Delta}_{\xi, xy} ^{-1}\right)^T &= \cc \tilde{\mathds{1}}\sum _{n=0} ^{\infty} \mathbf{S}_{x_n,y}^T \left(\prod _{\substack{i=1 \\ n>0}} ^{n}   \left(-\mathbf{X}_{\xi \xi, x_i}\right)^T \mathbf{S}_{x_{i-1}x_i}^T \right)  \\
&= \sum _{n=0} ^{\infty} \mathbf{S}_{y x_n} \tilde{\mathds{1}}\cc \left(\prod _{\substack{i=1 \\ n>0}} ^{n}  \left(-\mathbf{X}_{\xi \xi, x_i}\right)^T \mathbf{S}_{x_{i-1} x_i}^T \right) \\
&= \sum _{n=0} ^{\infty} \mathbf{S}_{y x_n} \tilde{\mathds{1}}\cc \left(\prod _{\substack{i=1 \\ n>0}} ^{n} \left(-\mathbf{X}_{\xi \xi, x_i}\right)^T \tilde{\mathds{1}}\cc ^{-1} \tilde{\mathds{1}}\cc  \mathbf{S}_{x_{i-1} x_i}^T \tilde{\mathds{1}}\cc ^{-1} \tilde{\mathds{1}}\cc \right)  \\
&= \sum _{n=0} ^{\infty} \mathbf{S}_{y x_n} \tilde{\mathds{1}}\cc \left(\prod _{\substack{i=1 \\ n>0}} ^{n}    \left(-\mathbf{X}_{\xi \xi, x_i}\right)^T \tilde{\mathds{1}}\cc ^{-1}   \mathbf{S}_{x_i x_{i-1}}  \tilde{\mathds{1}}\cc \right)  \\
&= -\sum _{n=0} ^{\infty} \mathbf{S}_{y x_n}  \left(\prod _{\substack{i=1 \\ n>0}} ^{n} \left(-\mathbf{X}_{\xi \xi, x_i}\right)    \mathbf{S}_{x_i x_{i-1}} \right)  \tilde{\mathds{1}}\cc ^{-1}  \\ 
&= -\mathbf{\Delta}_{\xi yx} ^{-1},
\end{align}
where we used that 
\begin{align}
\cc \tilde{\mathds{1}} \mathbf{X}^T_{\xi \xi} \tilde{\mathds{1}} \cc^{-1}=\mathbf{X}_{\xi \xi}.
\end{align}
Noting that
\begin{align}
\tilde{\mathbf{X}}^T_{\xi \Xi}&=-\tilde{\mathbf{X}}_{\Xi \xi}, \\
\tilde{\mathbf{X}}^T_{\xi \Phi}&=\tilde{\mathbf{X}}_{\Phi \xi}, \\
\tilde{\mathbf{X}}^T_{\xi \phi}&=\tilde{\mathbf{X}}_{\phi \xi},
\end{align}
the validity of \eqref{eq:shiftCondition} follows immediately.
%
%\begin{align}
%\mathbf{S}(x,y)(-i\overleftarrow{\slashed{\partial}^y}-M_\xi-\mathbf{A}(y))&=\delta (x-y)+\sum _{k=1} ^{\infty} \left(\prod_{i=1} ^k %%\int \rd^d x_i \; \mathbf{S}_f(x_{i-1}-x_i)\mathbf{A}(x_i) \delta(x_k-y) \right) \nonumber \\ & \quad  -\sum _{k=0} ^{\infty} \left(\prod %_{\substack{i=1 \\ k>0}} ^{k} \int \rd^d x_i \; \mathbf{S}_f(x_{i-1}-x_i) \mathbf{A}(x_i)\right) \mathbf{S}_f(x_k-y) \mathbf{A}(y)=%\delta(x-y)
%\end{align}

\section{Loop functions}
\label{sec:loop_functions}
The integrals $\ZI [q^{2n_c}]^{n_i n_j \dots n_L} _{i j \dots 0}$ are defined as in \cite{Zhang:2016pja}, that is 
\begin{align}
  \int \frac{\rd^d q}{(2\pi)^d} \frac{q^{\mu_1} q^{\mu_2} \dots q^{\mu_{2n_c}}}{(q^2-M_i^2)^{n_i}(q^2-M_j^2)^{n_j}\dots (q^2)^{n_L}}\equiv \frac{i}{16 \pi ^2} g^{\mu_1 \mu_2 \dots \mu_{2n_c}} \ZI [q^{2n_c}]^{n_i n_j \dots n_L} _{i j \dots 0},
\end{align}
where $g^{\mu_1 \mu_2 \dots \mu_{2n_c}}$ is the completely symmetric
combination of metric tensors with $2n_c$ indices, for instance
$g^{\mu \nu \rho \sigma}=g^{\mu \nu}g^{\rho \sigma}+g^{\mu \rho}g^{\nu
  \sigma}+g^{\mu \sigma}g^{\nu \rho}$.  For $n_c=0$ we define the
shorthand notation
$\ZI[q^0]^{n_i n_j \dots n_L} _{i j \dots 0} \equiv \ZI^{n_i n_j \dots
  n_L} _{i j \dots 0}$.  The integrals can be reduced to basis
integrals using the reduction relations \cite{Zhang:2016pja}
\begin{align}
  \ZI[q^{2n_c}] ^{n_i n_j \dots n_L} _{i j \dots 0} &=
  \frac{1}{\Delta^2 _{ij}}\left(\ZI[q^{2n_c}]^{n_i n_j-1 \dots n_L}-\ZI[q^{2n_c}]^{n_i-1 \, n_j \dots n_L}\right), \\
  \ZI[q^{2n_c}] ^{n_i n_j \dots n_L} _{i j \dots 0} &=
  \frac{1}{M_i^2}\left(\ZI[q^{2n_c}]^{n_i n_j \dots n_L-1}-\ZI[q^{2n_c}]^{n_i-1 \, n_j \dots n_L}\right),
\end{align}
where $\Delta^2 _{ij}=M_i^2-M_j^2$.
For convenience we have included the reduction algorithm and the basis
integrals in the ancillary Mathematica file \texttt{LoopFunctions.m}
in the arXiv submission of this publication with the correspondence
\begin{align}
  \ZI [q^{2n_c}]^{n_i n_j \dots n_L} _{i j \dots 0} \equiv
  J[n_c, \{\{M_i,n_i\}, \{M_j,n_j\}, \ldots\}, n_L] .
\end{align}

\section{Useful relations for spinors and $SU(N)$ groups}
\label{sec: spinor algebra}
We define the charge conjugate $\ccfield{\psi}$ of a 4-spinor $\psi$
as
\begin{align}
  \ccfield{\psi} &\equiv \cc \bar{\psi}^T, &
  \overline{\ccfield{\psi}} &= \psi^T \cc,
\end{align}
where $\cc$ is the charge conjugation operator and
$\bar{\psi}=\psi^\dagger \gamma^0$. It follows from this definition
that
\begin{align}
  \ccfield{(\psi_R)} &= \cc\,\overline{\psi_L}^T, &
  \ccfield{(\psi_L)} &= \cc\,\overline{\psi_R}^T.
\end{align}
The following properties of $\cc$ hold in the Dirac and Weyl
representation:
\begin{align}
  \cc &= i\gamma^2\gamma^0, \\
  \cc &= -\cc^{-1} = -\cc^\dagger = -\cc^T, \\
  \cc \gamma^\mu \cc^{-1} &= - (\gamma^\mu)^T, \\
  \cc \gamma^5 \cc^{-1} &= (\gamma^5)^T = \gamma^5, \\
  \cc \gamma^5 \gamma^\mu \cc^{-1} &= (\gamma^5 \gamma^\mu)^T = (\gamma^\mu)^T \gamma^5, \\
  \cc P_L \cc^{-1} &= (P_L)^T = P_L, \\
  \cc P_R \cc^{-1} &= (P_R)^T = P_R.
\end{align}
In our formalism we require that if a model contains Dirac spinors
$\psi$, then the Lagrangian is expressed in terms of $\psi$ and
$\bar{\psi}$.  If the model contains Majorana spinors $\lambda$, we
require that the Lagrangian is expressed \emph{only} in terms of
$\lambda$, but \emph{not} in terms of $\bar{\lambda}$.  Note that
$\bar{\lambda}$ can always be rewritten as
\begin{align}
  \bar{\lambda} &= (\ccfield{\lambda})^T \cc = \lambda^T \cc
\end{align}
because for Majorana fermions $\ccfield{\lambda} = \lambda$.  When
contracting spinor indices the following identity may be used
\begin{align}
  \psi^T \Gamma^T \bar{\psi}^T &= - \bar{\psi} \Gamma \psi.
\end{align}
A useful relation for the generators $T^a$ of the fundamental representation of $SU(N)$ is
\begin{align}
T^a_{ij}T^a_{kl}=\frac{1}{2}\left(\delta_{il}\delta_{jk}-\frac{1}{N}\delta_{ij}\delta_{kl}\right).
\label{eq:genrel}
\end{align}

\section{Dimensional regularization and dimensional reduction}
\label{sec:DREG_DRED}

Throughout this publication we have assumed that the models are
regularized in dimensional regularization (DREG)
\cite{tHooft:1972tcz}, where loop calculations are performed in a
quasi-$d$-dimensional space $QdS$ with the metric tensor
$\ddim{g}^{\mu\nu}$ with the property
\begin{align}
  \ddim{g}^{\mu\nu} \ddim{g}_{\mu\nu} &= d = 4 - \eps.
\end{align}
Although DREG is suited for non-supersymmetric models, it is
cumbersome to use in supersymmetric models, as it explicitly breaks
supersymmetry \cite{Delbourgo:1974az}.  For supersymmetric models
regularization by dimensional reduction (DRED) \cite{Siegel:1979wq} is
more suited, because it is currently known to not break supersymmetry
up to the three-loop level
\cite{Capper:1979ns,Stockinger:2005gx,Stockinger:2018oxe}.
In DRED the quasi-$4$-dimensional space, denoted as $Q4S$, is
decomposed into a quasi-$d$-dimensional space $QdS$ and a
quasi-$\eps$-dimensional space $Q\eps S$, as
$Q4S=QdS\oplus Q\eps S$ \cite{Stockinger:2005gx}.  The
corresponding $4$- and $\eps$-dimensional metrics are denoted as
$\fourdim{g}^{\mu\nu}$ and $\epsdim{g}^{\mu\nu}$, respectively, and
the following properties hold:
\begin{align}
  \fourdim{g}^{\mu\nu} &= \ddim{g}^{\mu\nu} + \epsdim{g}^{\mu\nu}, \\
  \epsdim{g}^{\mu}_\sigma \fourdim{g}^{\sigma\nu} &= \epsdim{g}^{\mu\nu}, \\
  \ddim{g}^{\mu}_\sigma \fourdim{g}^{\sigma\nu} &= \ddim{g}^{\mu\nu}, \\
  \fourdim{g}^{\mu\nu} \fourdim{g}_{\mu\nu} &= 4, \\
  \ddim{g}^{\mu\nu} \ddim{g}_{\mu\nu} &= d, \\
  \epsdim{g}^{\mu\nu} \epsdim{g}_{\mu\nu} &= \eps, \\
  \epsdim{g}^{\mu\nu} \ddim{g}_{\mu\nu} &= 0, \\
  \tr(\gamma^\mu\gamma_\mu) &= 4d.
\end{align}

%%%%%%%%%%%%%%%%%%%%%%%%%%%%%%%%%%%%%%%%%%%%%%%%%%

\bibliographystyle{JHEP}
\bibliography{paper}

\end{document}